# Bio-integrated μBots with Overtone Ultra-Wideband Magnetoelectric Antennas for Wireless Telemetry

Mahdieh Shojaei Baghini[1], Adam Armada-Moreira[2], Alessio Di Clemente[2,3], Dibyajyoti Mukherjee[1], Afesomeh Ofiare[4], Jonathon Harwell[5], Mary Dysko[6], Luana Benetti[8], Declan Bolster[9], Laura Mazon Maldonado[1], Dayhim Nekoeian[1], Moreno Maini[7], Mostafa Elsayed[1], Rossana Cecchi[2], Ricardo Ferreira[8], Jeff Kettle[5], Sandy Cochran[6], William Holmes[9], Carlos Garcia Nunez[1], Luca Selmi[7], Nicola Toschi[10,11], Michele Giugliano[2,3,12], Hadi Heidari[1]

[1] Microelectronics Laboratory, James Watt School of Engineering, University of Glasgow, G128QQ, United Kingdom

[2] Department of Biomedical, Metabolic and Neural Sciences, University of Modena and Reggio Emilia, Via Campi 287, 41125 Modena, Italy

[3] International School for Advanced Studies (SISSA), Via Bonomea 265, 34136 Trieste, Italy

[4] Centre for Advanced Electronics, University of Glasgow, G128QQ, United Kingdom

[5] James Watt School of Engineering, University of Glasgow, G128QQ, United Kingdom

[6] Centre for Medical & Industrial Ultrasonics, James Watt School of Engineering, University of Glasgow, Glasgow G12 8QQ, United Kingdom

[7] Department of Engineering "Enzo Ferrari", via Pietro Vivarelli 10, University of Modena and Reggio Emilia, 41125 Modena, Italy

[8] International Iberian Nanotechnology Laboratory (INL), Braga, Portugal

[9] School of Psychology and Neuroscience, College of Medicine, Veterinary and Life Science, University of Glasgow, Glasgow G61 1QH, United Kingdom

[10] Department of Biomedicine and Prevention, University of Rome, "Tor Vergata", Rome, Italy

[11] Department of Radiology, Athinoula A. Martinos Center for Biomedical Imaging, Massachusetts General Hospital, Harvard Medical School, Charlestown, MA, United States

[12] National Interuniversity Consortium of Materials Science and Technology (INSTM), Florence, Italy

*Corresponding authors: Mahdieh Shojaei Baghini, Adam Armada-Moreira, Hadi Heidari*

*Abstract*

Implantable and wearable devices require antennas that are both miniaturized and efficient, yet conventional designs are constrained by narrow bandwidth and orientation sensitivity. We report overtone ultra-wideband magnetoelectric (OUWB-ME) antennas that exploit higher order acoustic modes in polished silicon substrates to achieve a 22.6 GHz bandwidth in the 3-4 GHz range. Packaged into "µBots," these magnetoelectric heterostructures bonded with silver nanoparticle inks maintain stable operation under biological loading. *In vitro* assays confirm the biocompatibility of AlN and the protective role of parylene encapsulation for FeGa. *Ex vivo* rat and human tissues reshape transmission spectra, identifying reproducible frequency windows near 3.3 and 3.9 GHz. µBots enable real-time audiovisual telemetry using software-defined radios and exhibit compatibility with 7T MRI. By combining wideband response, robustness to misalignment, and biocompatible packaging, OUWB-ME µBots provide a scalable platform for wireless bio-integrated communication and telemetry.

# Introduction

In recent years, advances in wireless communication and antenna design for wearable and implantable bioelectronics have enabled sophisticated, real-time health monitoring, deep-tissue neuromodulation and minimally invasive therapeutics, opening new routes to earlier diagnosis of neurodegenerative diseases (*1-4*). Early bioelectronic interfaces relied on tethered probes that breach tissue barriers and heighten infection risk (*5*). These limitations drove a shift to wireless energy and signal delivery, including inductive coupling (*6-8*), acoustic power transfer (*9, 10*), optogenetic control (*11, 12*), and electromagnetic (EM) antennas (*13, 14*). Inductive links are widely explored, offering simple architectures and efficient near-field power or data transfer. However, their range is intrinsically short, and coil angular misalignment severely degrades performance, leading to substantial signal loss (*15*). Because antenna power-transfer efficiency scales with size, aggressive miniaturization degrades performance and complicates implant integration. By contrast, acoustically coupled systems use compact transducers that integrate readily with bioelectronics and have been successfully incorporated into diverse biomedical devices (*9*). Compared with inductive links, acoustic links are less sensitive to misalignment but strongly dependent on the coupling medium. Acoustic links incur frequency-dependent attenuation, where losses in bone and soft tissue can be $\geq 22$ dB cm$^{-1}$ MHz$^{-1}$ (*10*). Optical/optogenetic channels ease alignment constraints via line-of-sight transmission and have been demonstrated in several bioelectronic platforms (*16*), but they require external power or batteries and are restricted to unobstructed paths. By contrast, radio frequency (RF)/EM antennas operate in both near- and far-field regimes and support aggressive miniaturization, improving suitability for implantable and wearable systems (*14, 17*). Although RF antennas can be miniaturized, particularly at multi-GHz frequencies, practical designs rarely shrink beyond $\sim \lambda/10$ (*18*), where $\lambda$ is the signal wavelength. As size scales inversely with frequency, pushing higher in the spectrum reduces footprint but worsens dielectric losses in soft tissue and bone, elevating specific absorption rate (SAR) and thermal load during prolonged operation. These trade-offs motivate alternative modalities for deep-body implants that are safe for chronic use. Recently, magnetically driven magnetoelectric (ME) antennas that harness acoustic (mechanical) resonance have gained considerable attention. By operating at acoustic rather than electromagnetic resonance, they permit miniaturization by several orders of magnitude relative to conventional RF antennas (*19-21*). ME antennas transduce magnetic fields via a magnetostrictive layer that strains an underlying piezoelectric layer/material, producing an electrical output (*19, 22*). Most reported devices, laminated heterostructures such as lead zirconate titanate (PZT)/Metglas, lead magnesium niobate-lead titanate/Metglas, polyvinylidene fluoride/Metglas and PZT/Terfenol-D, operate below ~300 kHz and occupy millimeter-scale footprints (*5, 19, 23*). Low-frequency ME antennas are widely used for short-range links in the order of 1-5 cm and wireless power harvesting, and stimulation in implantable systems (*24-26*). By contrast, high-frequency silicon-based ME platforms enable deeper miniaturization and CMOS co-integration. Microfabricated ME resonators, thin-film bulk acoustic (FBAR) and solidly mounted resonators (SMR) produced via bulk or surface micromachining, typically operate from the low-MHz to GHz regime (*27-31*). However, the designs reported thus far exhibit low operational bandwidth (< 10 MHz) and rely on the presence of an air-gap or Bragg reflectors to obtain fundamental single or dual resonance modes of operation. Higher-order acoustic modes, referred to as overtones, emerge when acoustic resonators sustain standing waves beyond the fundamental thickness-extensional resonance and provide a promising exploitation route towards the development of planar and implantable magnetoelectric antennas. Their appearance is strongly influenced by material choice and substrate quality, with aluminium nitride, zinc oxide, and lithium niobate being particularly prone due to strong piezoelectric coupling, and low-stress, double-side polished silicon substrates further amplifying these effects through efficient acoustic reflection at smooth boundaries (*32*).

In this work we present overtone ultra-wideband (OUWB) ME antenna, packaged by precision printing of silver-nanoparticle bonds (Fig. 1A-B). Rather than suppressing artefacts, OUWB-ME antennas deliberately harness typically undesirable, yet often unavoidable, overtone reflections arising in low-stress polished mechanical substrates and magnetoelectric thin-film stiffness mismatch (Fig. 1C) leading to an ultrawide 22.6 GHz bandwidth, the largest reported to date. Using double sided polished silicon, which exhibits a lowered residual stress and supports high-order acoustic overtones, we leverage the electromechanical coupling of BCC $Fe_{0.79}Ga_{0.21}$ and wurtzite AlN to access modes in the 3-4 GHz range. Operating in the 3-4 GHz band strikes a pragmatic balance between physics and physiology: wavelengths short enough for millimeter-scale, efficient antennas, yet long enough to maintain usable penetration through soft tissue. Compared with crowded sub-3 GHz ISM (Industrial, Scientific, and Medical) allocations, it offers wider contiguous bandwidth and a cleaner interference environment, enabling higher-throughput links for implants and wearables (*33*).

The packaged OUWB-ME antennas, herewith referred to as "µBots", are interfaced with commercial transceivers to enable audio-visual transmission of sonogram data (Fig. 1D). Given their metallic and ferromagnetic constituents, we assess essential MRI compatibility at 7T. We further conduct systematic *in vitro* and *ex vivo* studies in rat and human tissues to confirm biocompatibility and quantify how tissue modulates transmission and overtone strength. Collectively, the analyses provide necessary and previously unevaluated results towards the design and deployment of magnetoelectric antennas for wireless power and data transfer. Harnessing the standing wave conditions that excite higher-order overtones beyond the fundamental mode, an intrinsic large bandwidth emerges, accompanied by the coexistence of multiple bands that extend communication capabilities beyond conventional resonance modes.

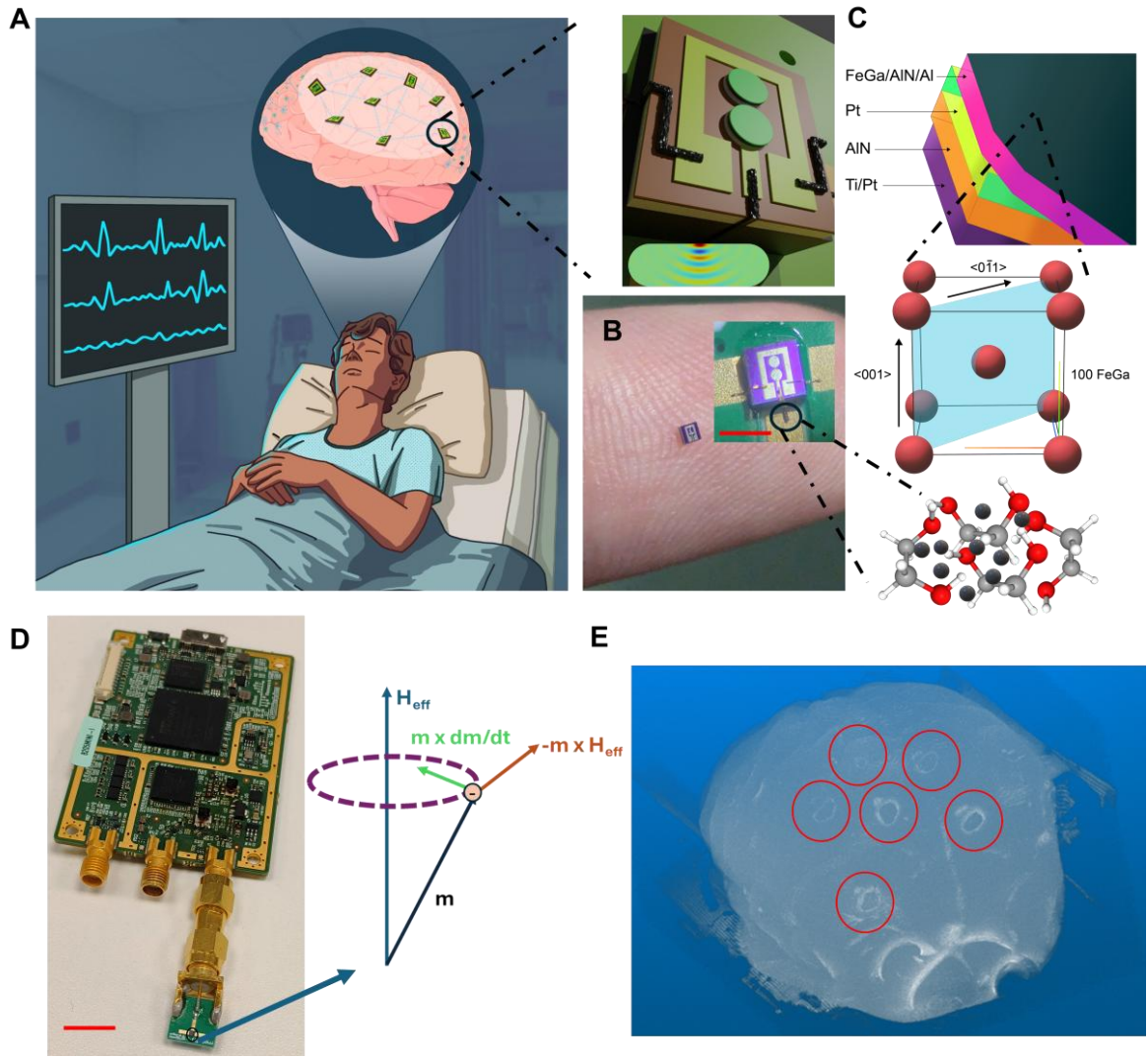

**Fig. 1. (A)** μBot federation wireless platform for bio-integrable communication consisting of packaged overtone ultrawideband (OUWB) magnetoelectric (ME) antennas with active area dominated by the circular components in green. Concept: a distributed network of ME nodes embedded in tissue enabling wireless sensing/telemetry to an external receiver (inset: acoustic wave pathway in the polished substrate with lowered stress mounting the OUWB-ME antenna). **(B)** Microfabricated OUWB-ME antenna at the fingertip scale and mounted on a ceramic PCB (inset, scale bar: 1 mm) via ethylene glycol-based silver nanoparticle inks capable of RF transmission. **(C)** Device architecture consisting of AlN/$Fe_{0.79}Ga_{0.21}$ (110 BCC) engineered to stiffness mismatch of 1:4.82 to enhance bandwidth of antenna for audio-visual transmission of signals while miniaturizing size. **(D)** Packaged OUWB-ME antenna (μBot) connected to Universal Software Radio Peripheral (USRP) platform (scale bar: 1 cm). The precession of particles, under an external magnetic field $H_{eff}$ in the ferromagnetic layer is influenced by the conservative and dissipative magnetic torques of the magnetostrictive thin film. **(E)** Volumetric image projection post 7T-MRI of 6 μBots implanted in Agar.

# Results

## Design and Characterization of OUWB-ME Antennas and µBots

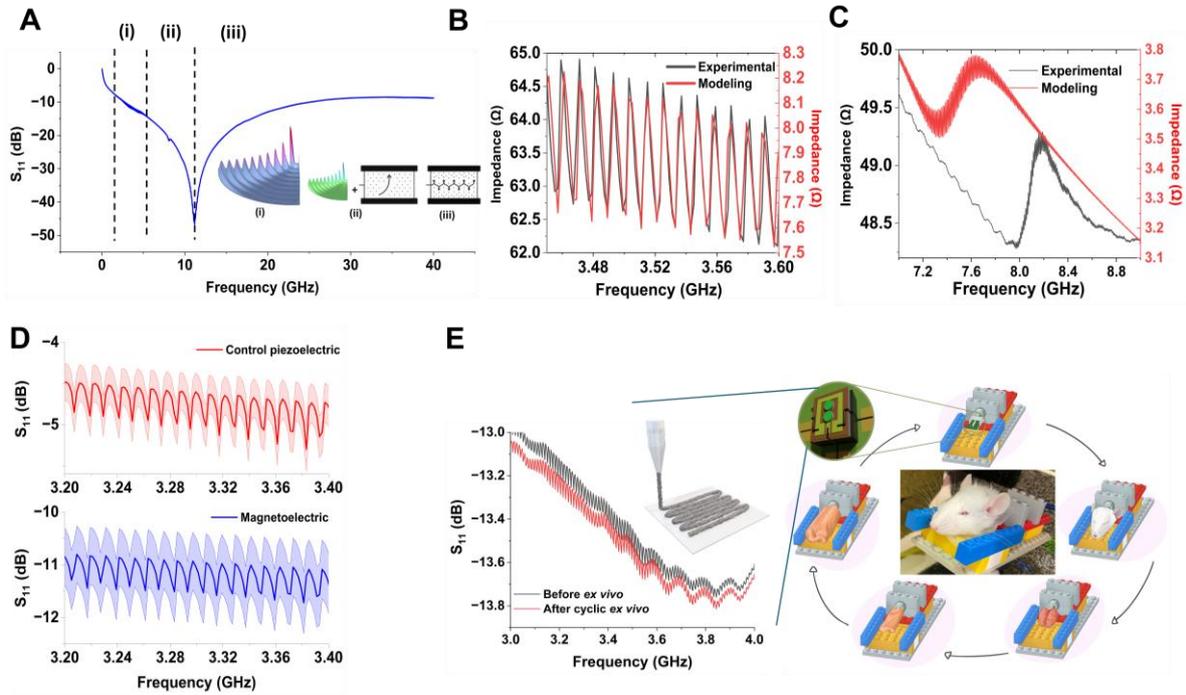

**Fig. 2. (A)** Ultrawideband reflection parameter of OUWB-ME antennas operating primarily in (i) overtone acoustic region, (ii) higher-order acoustic harmonic(s) and onset of self-resonance dominated by parasitic inductance and (iii) self-resonance wherein the electromagnetic field transmission dominates acoustic modes and is largely dominated by piezoelectric layer thickness. **(B)-(C)** Experimental and modeled impedance spectra in the overtone and higher order harmonic region of the OUWB-ME antennas that are utilized for extraction of acoustic properties of piezoelectric and magnetostrictive materials in the RF regime via the frequency and dynamic range. The absolute value of the experimental impedance is prone to manual artifacts such as probe landing and bond quality leading to mismatch with theoretical results in magnitude. **(D)** Comparison of reflection parameter between piezoelectric control resonators and OUWB-ME antennas highlighting the improvement in return loss. Shaded band represents ± 5% uncertainty error primarily due to calibration. the trace is apparently thicker due to reproducible small oscillations of the trace as seen in Fig. 2B and Fig. 2D **(E)** Reflection parameter of packaged OUWB-ME antennas (µBots) with nanoparticle-based inks prior to and after cyclic *ex vivo* testing under severed rat head, explanted rat brain, and human cortical autoptic tissue highlighting bonding stability.

The OUWB-ME antennas consist of 453 nm of AlN and $Fe_{0.79}Ga_{0.21}$ [(25 nm)/AlN(5 nm)]$_7$ as the active transduction layer along with a capping layer of 5 nm of Al exhibiting an active area of 0.038 mm$^2$ with an ultrawide bandwidth of 22.688 ± 0.256 GHz. The wideband characteristics of the devices can be divided into three distinct regions, (i) the primary overtone region, (ii) the spurious overtone and self-resonance onset followed by (iii) self-resonance as shown in Fig. 2A. Engineering a frequency separation between the magnetoelectric overtone resonances and the self-resonance increases transmission efficiency, mitigating the effects of the piezoelectric layer's finite impedance. Furthermore, in the overtone region, application of an in-situ DC magnetic field enabled deterministic modulation of the OUWB-ME antenna transmission spectra across devices (Fig. S1A-B). The spatial variation across a single wafer, set by the static magnetic field applied during deposition, provides a practical route to fine-tune OUWB-ME antennas without altering geometry. As seen in Fig. S1A, in the first overtone, the differential transmission metric rises from ~ 0 dB at the low-field center (device position 8) to 1.23-1.47 dB toward one side (positions 4-5) and reaches 2.63 dB at the opposite edge (position 10). The 2.63 dB extremum corresponds to ≈ 1.8× higher transmitted relative to the center, yielding a wafer-intrinsic "gain ladder." In practice, devices can be binned straight from the same sample, ~ 0 dB parts for nominal links, 1-1.5 dB parts to equalize arrays within ± 0.5 dB, and 2.63 dB "hot" parts where margin or misalignment tolerance is most demanding. Because this trimming derives from magnetization order rather than resonant geometry, it preserves the device's wideband character, enabling deterministic, fabrication-level amplitude setting without re-matching. This magnetic control fosters the assembly of OUWB-ME antenna federations with programmable transmission power within a single fabrication run, thereby advancing scalable deployment. Integration of the spin-orbit-coupled layer, together with the dynamic terminating impedance imparted by the polished substrate, suppresses reflections and, in turn, broadens the operational bandwidth of the OUWB-ME antennas. The deliberate incorporation of a lossy yet low-stress substrate increases acoustic damping, thereby reducing antenna insertion loss. While techniques such as piezoresponse force microscopy and the Berlincourt method facilitate extraction of quasi-static material characteristics, we harness the overtone region to additionally resolve the RF properties of the materials via a modified multi-variable Sittig's method, a critical factor in the design of OUWB-ME antennas and in predicting both their fundamental overtones and higher-order harmonics (Fig. 2B-C, Materials and Methods, Table S1). The ferromagnetic crystal was optimized during sputtering towards the development of anisotropic (110) BCC $Fe_{0.79}Ga_{0.21}$, as confirmed by X-Ray Diffractometry (Fig. S2), attributed to ordered D03 and unordered A2 mixed phases which has been shown to enhance static magnetostriction (34). The CMOS-compatible piezoelectric AlN exhibits a c-axis formation of grain size of 167.8 nm. The confounding effects of symmetry mismatch and limited adatom mobility led to increased grain boundaries, yielding small, misoriented columnar grains, rough or partially amorphous interfaces, and residual-stress gradients that boost scattering and internal friction, contributing to the high bandwidth of the antennas (Fig. S3). In addition, the strong modulus mismatch ($Fe_{0.79}Ga_{0.21}$ = 59.95 GPa versus AlN = 289 GPa; Materials and Methods) concentrates misfit/thermal strain within the BCC layer, elevates interfacial shear on AlN, and visco-elastically loads the piezoelectric, enhancing acoustic impedance contrast and damping. This effect is accentuated in the reflection spectra; relative to piezoelectric controls, OUWB-ME antennas show more than a twofold suppression in the overtone region, as shown in Fig. 2D, yielding wider bandwidth while enabling magnetostrictive control. The net effect is further bandwidth broadening at the expense of Q-factor reduction (35) (Fig. S4). Magnetometry measurements confirm a decrease in intrinsic coercivity and identify the hard axis as perpendicular to the sample plane, as indicated by a nearly hysteresis-free magnetization loop (Fig. S5). To improve implantability of the OUWB-ME antennas, bonding was carried out via an ethylene-glycol-based high viscosity silver conductive ink (4.2 μΩ cm) under micro-dispensing. Owing to dielectric losses in both the PCB and the nanoparticle ink, the reflection parameter increased yet stayed below -10 dB throughout the overtone region as shown in Fig. 2E, with stable performance over repeated *ex vivo* cycles. Phase shifts and parasitic resonances introduced by the metallic nanoparticle ink-bonded interfaces occur outside the primary magnetoelectric overtone band (3.0-3.6 GHz) and therefore do not perturb the OUWB ME antenna-µBot response within the band of interest (Fig. S6).

*In vitro biocompatibility analysis of materials constituting the OUWB-ME Antenna*

The active materials that compose the OUWB-ME antenna, in addition to Parylene, were used as substrates for primary cortical neuronal cultures, to ascertain their individual biocompatibility. Several biocompatibility aspects were targeted: cell viability, neuronal-glial balance within the culture, and neuronal function. Regarding cell viability, the level of mitochondrial activity was measured weekly over a month, using a colorimetric assay that relies on cellular metabolism, generating a colored formazan compound when exposed to NADH/NADPH (*36*). Thus, following incubation, the absorbance at 450 nm is a measure of cellular metabolism, acting as an indirect quantification of neuronal health. While AlN presented levels of cell viability equivalent to those of the control condition (glass coverslips), cells growing on FeGa and AlN/FeGa/Parylene substrates show a significant decrease in cell viability (Fig. 3A). The reduction in mitochondrial activity of cells growing on the FeGa substrate is linked to low cell survival. Indeed, cells were unable to form a monolayer on this substrate and only a small portion of neurons were able to survive, surrounded by astrocytes presenting an enlarged morphology, as seen in the immunostaining micrographs (Fig. 3B). In the case of the AlN/FeGa/Parylene substrates, the reduced metabolic activity is likely attributable to alterations in cellular architecture induced by cell-material interactions. While AlN allows for the formation of a cellular monolayer like that existing on glass, the AlN/FeGa/Parylene substrate leads to cell aggregation (Fig. 3B). The shift from monolayer to cell aggregates indicates a lack of neuronal adherence to the parylene C substrate, as has been previously observed (*37*).

All samples were subsequently stained for immunocytochemistry quantification (Fig. 3B, bottom row). In this case, no significant differences were found in terms of cell density (Fig. 3C). Analysis of the astrocyte-to-neuron coverage ratio revealed a significant shift only in cultures on FeGa, characterized by a reduction in neuronal area and a predominance of astrocytic coverage. The fact that cell density is slightly reduced in cultures growing on FeGa substrate but the area covered by astrocytes is equivalent to that of the control condition further corroborates the presence of hypertrophic astrocytes. This morphological change is a hallmark of the reactive astrocytic phenotype, which is associated with cellular stress responses and altered neuron-glia communication (*38, 39*). Interestingly, although the cells cultured on AlN/FeGa/Parylene show a different network architecture, the balance between glial and neuronal populations is maintained and only a decrease in the area covered by neurons can be identified, which is to be expected given the striking change in architecture caused by cell aggregation.

Since the aggregates do not allow for a deeper functional analysis of the cellular network, only cells cultured on the AlN and FeGa samples were functionally probed, using a fluorescent calcium indicator as a proxy for their intrinsic electrical activity. Calcium transients were detected in all experimental conditions (Fig. 3D, SI Videos 1-4) but, while cells growing on AlN presented calcium transient duration and amplitude similar to those cultured on glass coverslips, cells cultured on FeGa displayed longer calcium transients with significantly lower amplitude (Fig. 3E), which aligns with the finding that mostly only astrocytes were present (*40*). In all cases, the frequency of the calcium transients was equivalent to those plated on glass, as measured by the inter-spike interval. Finally, cells plated on AlN/FeGa/Parylene also displayed calcium transients (SI Video 3), once more corroborating that their lower cell viability read-out and altered cellular architecture were due to lack of cellular adherence and not material toxicity, and reiterating the protective role of Parylene C encapsulation of materials such as FeGa, to be explored during future iterations.

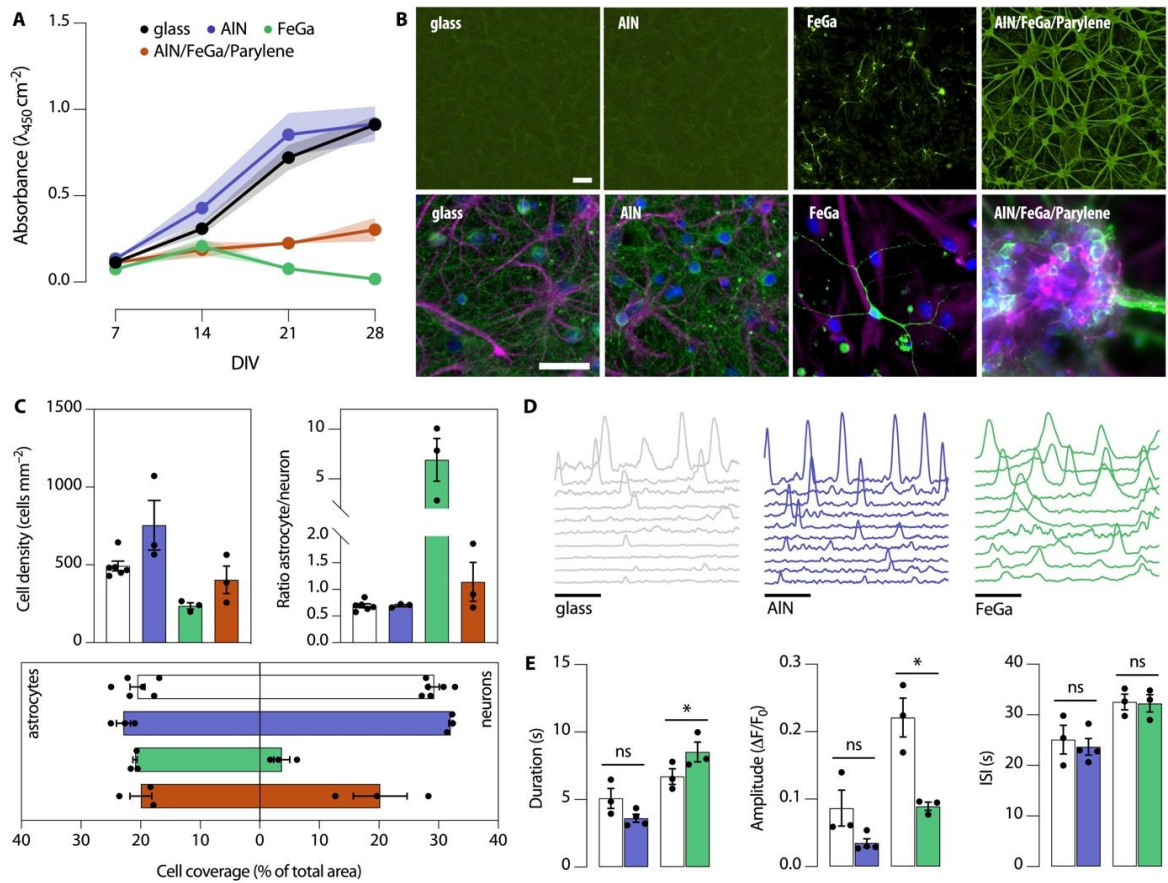

**Fig. 3. Assessment of material biocompatibility. (A)** Normalized cell viability, measured as absorbance at 450 nm, of cells plated on the different substrates. N = 3–6 independent cultures. ****p < 0.0001 in comparison with the control condition at the same DIV using a two-way ANOVA (days in vitro vs. substrate) followed by Tukey's multiple comparison test. **(B)** Representative fluorescent images of primary neuronal cultures plated on glass, and silicon wafers coated with AlN, FeGa, and AlN/FeGa/Parylene. On top row, $\beta_3$-Tubulin staining. On bottom row, neurons are represented in green ($\beta_3$-Tubulin), astrocytes in magenta (GFAP), and nuclei in blue (DAPI). Scale bar = 50 μm. **(C)** Quantitative analysis of cell density, ratio between neurons and astrocytes, and relative area covered by the different cell types. N = 3–6 independent cultures. ***p < 0.001 and ****p < 0.0001 in comparison with the control condition, using a one-way ANOVA followed by Dunnett's multiple comparison test. **(D)** Representative calcium signals measured from cells growing on glass, AlN, and FeGa. Scale bar = 30 s. **(E)** Quantitative analysis of calcium transient duration, amplitude, and frequency (represented as inter-spike interval, ISI). N = 3–4 independent cultures. *p < 0.05 using a paired t-test between experimental condition and respective control.

*Ex vivo transmission and reflection measurements of µBots*

Transmission ($S_{12}$) and reflection ($S_{22}$) parameters were measured using the µBots in conjunction with a commercially available horn antenna. To examine the influence of biological tissue on these parameters, various biological samples (rat and human), summarized in Table S2 and Fig. S7, were positioned directly over the µBots. Rat brain tissue was included alongside human samples to evaluate the µBot's response across different tissue types and conditions, including fresh tissue and intact head preparations. Testing across diverse tissues is essential, as differences in composition, water content, and structural organization influence acoustic and electromagnetic properties. An inert modeling clay of comparable weight to the rat head was included as a sham control for sample weight.

Regarding both $S_{12}$ (Fig. 4A) and $S_{22}$ (Fig. 4B) measurements, traces from each sample were normalized as fold-change relative to their respective blank µBots measurements. This normalization allowed for comparison across µBots despite baseline variability (Fig. S8). For reflection overtone analysis, the $S_{22}$ traces were detrended, allowing for the isolation of periodic components. Overtone strength was then calculated as the peak-to-peak amplitude of the denoised signal (Fig. 4C). Importantly, the existence of overtones in the measurement is caused by the µBots and not a consequence of electrical/measurement noise, since overtones are not present in open calibration measures (Fig. S9).

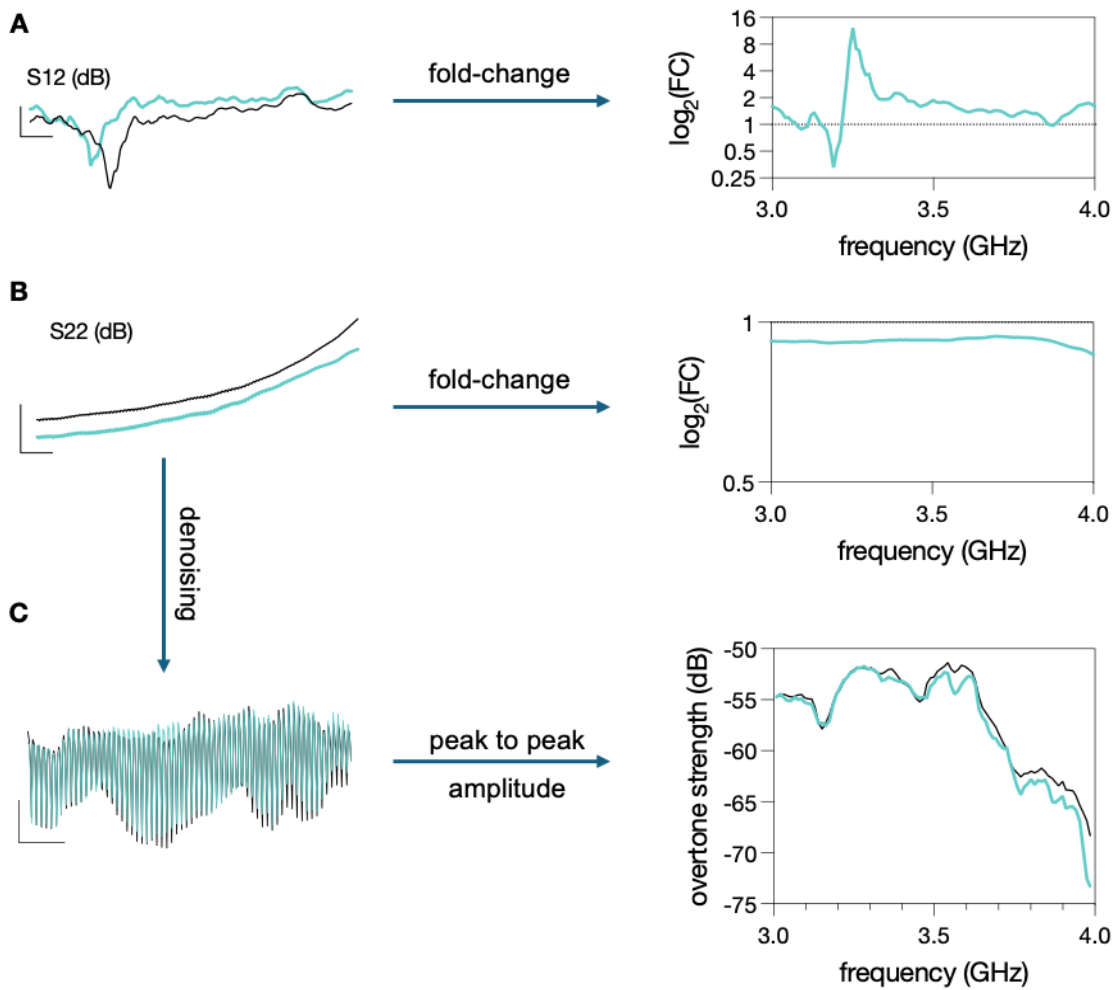

**Fig. 4. Experimental pipeline overview. (A)** Representative $S_{12}$ (transmission) measurement for blank µBot (black trace) and µBot with fresh rat brain (blue trace). X-scale bar = 0.1 GHz. Y-scale bar = 10 dB. The same biological condition measured in the different µBots were normalized in respect to their respective blank µBot measurement, generating a fold-change (FC) plot. **(B)** Representative $S_{22}$ (reflection) measurement for blank µBot (black trace) and µBot with fresh rat brain (blue trace). X-scale bar = 0.1 GHz. Y-scale bar = 2 dB. Similarly to $S_{12}$ data, sample measurements were normalized in relation to their respective blank antenna measurement, creating the FC dataset. **(C)** To quantify overtone strength, the $S_{22}$ signal was denoised. X-scale bar = 0.2 GHz. Y-scale bar = 0.001 absolute magnitude. From the denoised data, the overtone amplitude was calculated and used to generate the overtone strength plot.

Systematic analysis of the impact of biological tissue in µBot transmission (Fig. 5A and Fig. S10) showed that most samples exhibited transmission peaks at similar bandwidths (3.1, 3.27, and 3.95 GHz). Importantly, while the sham control also displayed an enhancement of transmission at these frequencies, the facilitation caused by biological tissue was generally stronger. Indeed, conditions involving fresh tissue (intact rat head, fresh brain, and cold brain, as well as unfixed human tissue) displayed a larger transmission peak around 3.1 GHz (Fig. 5B). This peak disappeared or decreased following tissue fixation, for rat and human samples respectively, whereas another distinct peak at 3.27 GHz (present under all conditions) was slightly accentuated in the fixed tissue (Fig. 5C). We hypothesize that fixation increases stiffness, imposing a fixed-constraint acoustic boundary, in the limiting case of a skull-intact rat head this produces a positive frequency shift, confirmed by overtone analysis (Fig. 7E), rather than the soft-boundary damping expected under compliant constraints. Since the acoustic resonance is responsible for magnetic field radiation, which is less impacted by the surface current, absorption in the soft tissue or bones indicates the optimal performance of the µBots at these frequencies. Finally, the peak found at 3.95 GHz followed the same pattern as that found at 3.1 GHz, with fresh tissue showing a larger increase when compared to their fixed counterparts (Fig. 5D). These shifts may additionally reflect changes in dielectric properties due to tissue fixation and dehydration on account of the varying dielectric constant in air and water (1 versus 80) leading to variable tissue coupling to the atmosphere (*41*) and constitute contributions in transmission arising from the parasitic effects of the conductive nanoparticle bonds, a previously unexamined but necessary component for the design of implantable magnetoelectric antennas. Furthermore, the cold brain exhibited a larger fold-change while maintaining the same dynamics, potentially due to the progressive warming and moving of the tissue during the measurement.

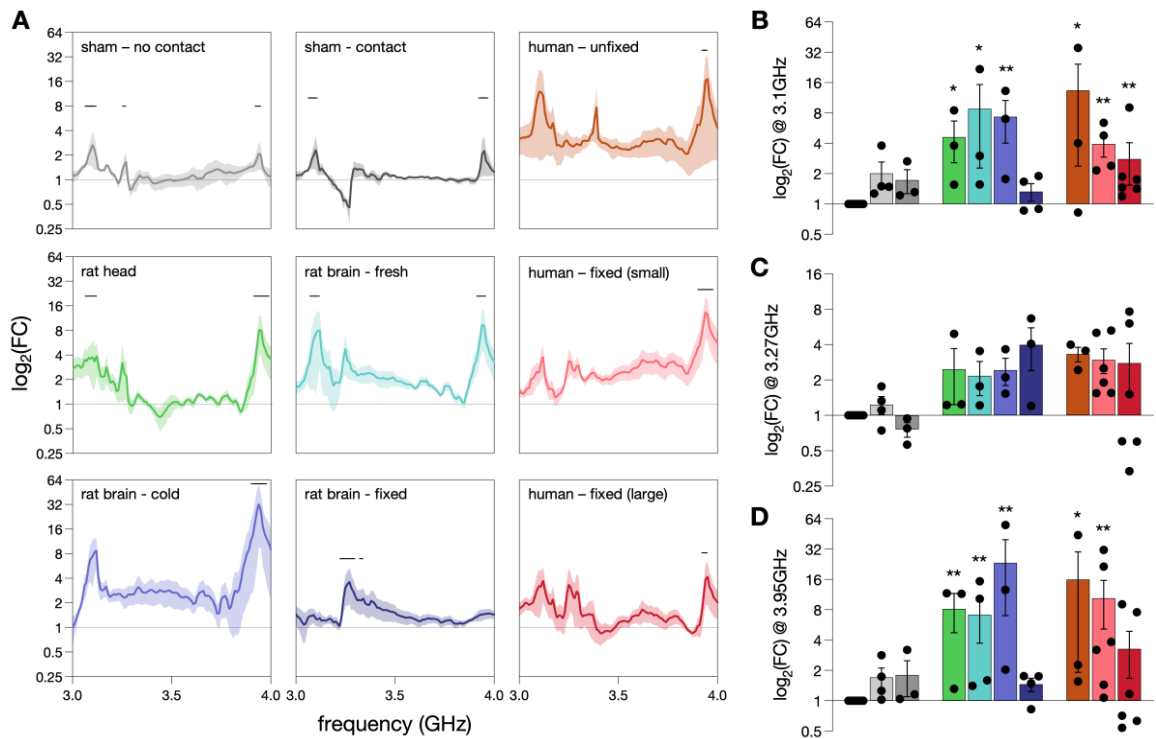

**Fig. 5. Impact of biological tissue on µBots transmission.** (**A**) $S_{12}$ fold-change (FC) for each biological condition. Data represented as mean ± SEM. N = 3–6 different µBots. Horizontal black lines represent significant differences between each condition and baseline. (**B**) Changes in $S_{12}$ at specific frequencies (3.1, 3.27, and 3.95 GHz). Data represented as mean ± SEM, with individual µBots measurements scattered as black dots. *$p < 0.05$, **$p < 0.01$ vs. baseline using Kruskal-Wallis test followed by the two-stage linear step-up procedure of Benjamini, Krieger and Yekutieli.

Reflection data (Fig. 6 and Fig. S11) revealed that the intact rat head, comprising brain, skull, and surrounding tissues, produced a noticeable damping effect but with a positive frequency shift, on account of increased stiffness. Hence, this was not exclusively due to sample weight, as tested using a similarly weighted sham control that showed no such effect. The other tested rat samples, including the isolated fresh brain, cold brain, and fixed brain, did not exhibit this damping, with the fresh brain slightly enhancing reflection. Among human samples, only the small fixed tissue showed a significant reflection attenuation, with the large fixed tissue trending similarly.

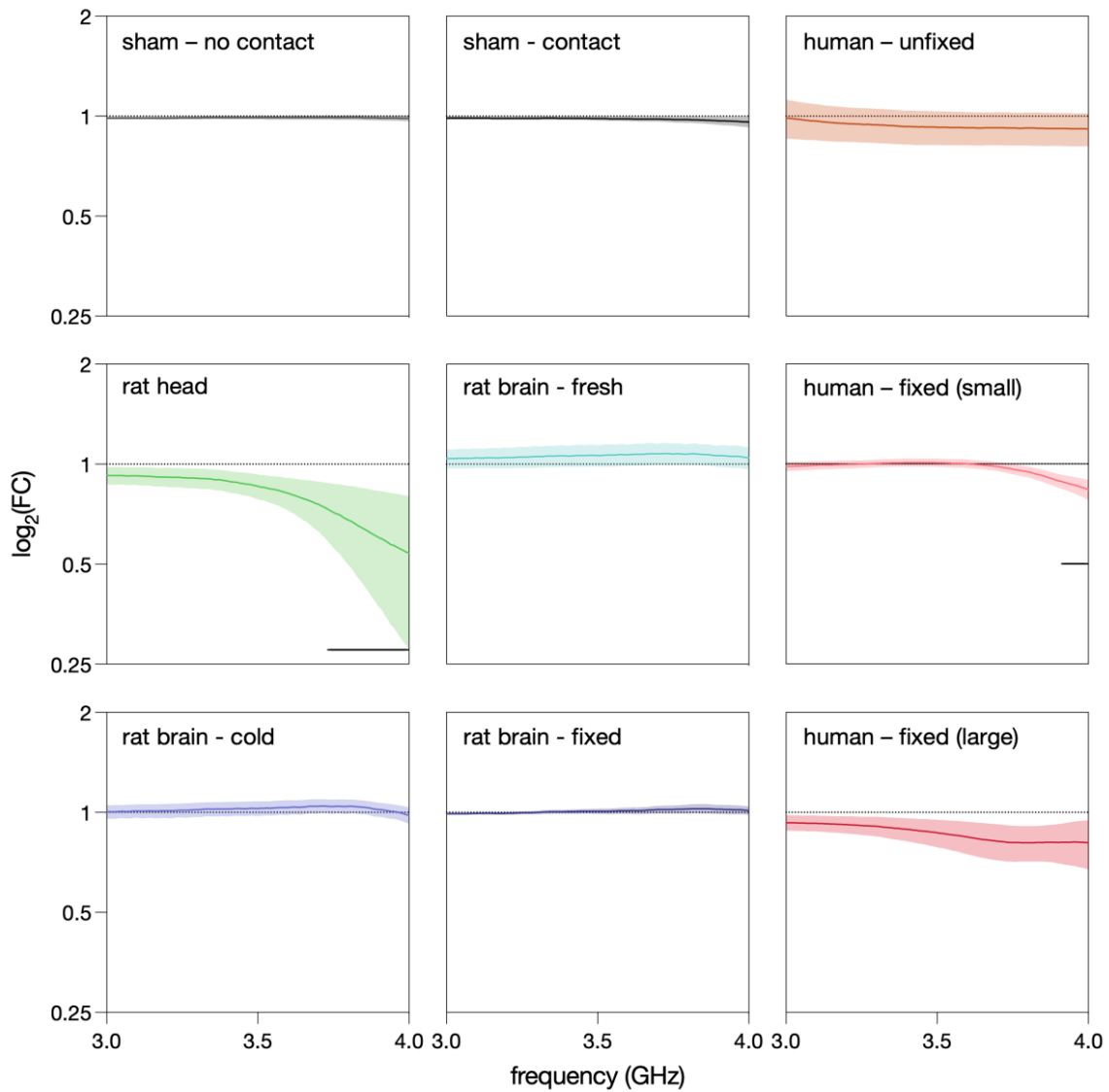

**Fig. 6. Impact of biological tissue on µBots reflection. (A)** $S_{22}$ fold-change (FC) for each biological condition. Data represented as mean ± SEM. N = 3–6 different µBots. Horizontal black lines represent significant differences between each condition and its control. All samples are mounted on non-conductive building blocks.

Detrended $S_{22}$ traces highlighted the presence of overtones across the ME range (Fig. S12). A first comparison between biological samples and their baseline suggested that the presence of the different samples could induce a frequency shift in the overtone strength signal dynamics (Fig. 7A). To confirm this observation, the optimal lag for each condition was identified using cross-correlation analysis (Fig. 7B), showing that the presence of the *ex vivo* severed rat head led to a shift toward higher frequencies, while the other biological samples showed a tendency to shift the signal toward lower frequencies. Importantly, the tested sham controls, in both contact and no-contact configurations, did not lead to frequency shifts.

Given that cross-correlation analysis can only quantify overall frequency shift, four main overtone features were defined in the overtone strength dynamics (Fig. 7C) - two maxima (P) and two minima (N) peaks, occurring sequentially (N1, P1, N2, P2) and present in all μBots at comparable frequencies and with similar strength (Fig. S13). Pooling all features for each tested condition revealed that only the severed rat head led to a significant shift toward higher frequencies, while the cold rat brain and all human samples displayed a significant negative frequency shift (Fig. 7Di). In terms of overtone attenuation caused by the different samples, the same trends were observed, albeit with increased inter-μBot variability (Fig. 7Dii). While all samples caused some degree of attenuation, only the explanted rat cold brain and the fixed small human tissue showed statistical significance. Noteworthily, the sham control in contact with the μBot also led to some overtone attenuation, which can be explained by its weight. The discrimination between weight per se and the impact of biological tissue will be explored below. Interestingly, the fresh rat brain did not impact overtone strength neither in frequency nor in attenuation.

Taken together, these results suggest that the severed rat head distinctly impacts overtone dynamics (Fig. 7E), likely reflecting the presence of skull and connective tissues. This emphasizes the importance of considering not only the target tissue but also the surrounding structures when assessing ME antenna performance in a biological environment. Moreover, the consistent negative frequency shift observed across all fixed samples suggests that fixation imposes common structural or dielectric changes, potentially linked to crosslinking of proteins and altered water mobility. The striking difference between explanted fresh and cold rat brain could stem from temperature- and fluid-related factors: the fresh brain remains warm and physiologically hydrated, while the cold brain contains additional aCSF and undergoes progressive warming during measurement, both of which affect its dielectric properties. The intermediate impact of the unfixed human tissue between these two conditions may also reflect a similar temperature- and water-content dependence. These observations point toward hydration state and tissue temperature, as well as the presence of surrounding tissue, as key modulators of overtone dynamics.

Since different tissues might have a differential effect on the different signal features, each overtone feature was specifically evaluated (Fig. 7F and Fig. S14). Despite having identical weight, the sham control and severed rat head displayed divergent patterns (Fig. 7Fi), where the sham did not preferentially affect any feature, whereas the rat head induced stronger frequency shifts and attenuation at higher frequency features (N2 and P2). Regarding explanted rat brain samples, the fresh brain did not differentially impact the features, while both the cold and fixed brain samples showed a pronounced effect on N1 (Fig. 7Fii) when compared to the other identified features. Human autoptic samples, though heavier overall, induced smaller frequency shifts and attenuation than the rat samples (Fig. 7Fiii). Also, these samples differentially affected N2, with more pronounced negative frequency shifts. Interestingly, the unfixed human autoptic sample did not attenuate N2. Finally, although the large fixed human sample was one order of magnitude heavier than its small counterpart, its signal attenuation was not proportionally larger. While these insights are preliminary, they reveal a relevant and unexplored influence of biological tissues on OUWB-ME antenna overtone dynamics. Most existing research on the interaction between implantable conventional radio frequency antennas and biological tissue focuses on detuning, attenuation, and power transfer efficiency, mostly relying on human body computational models or external skin contact only (*42-44*). Importantly, with this study, we were able to experimentally observe higher frequency shift induced by skull proximity and lower frequency shift in intracranially implantable magnetoelectric antennas due to tissue proximity. Previous studies only encompass traditional RF antennas in a theoretical study using computational human body models, albeit with a focus on general RF antenna detuning and not on specific ME antenna acoustic overtone dynamics (*43*).

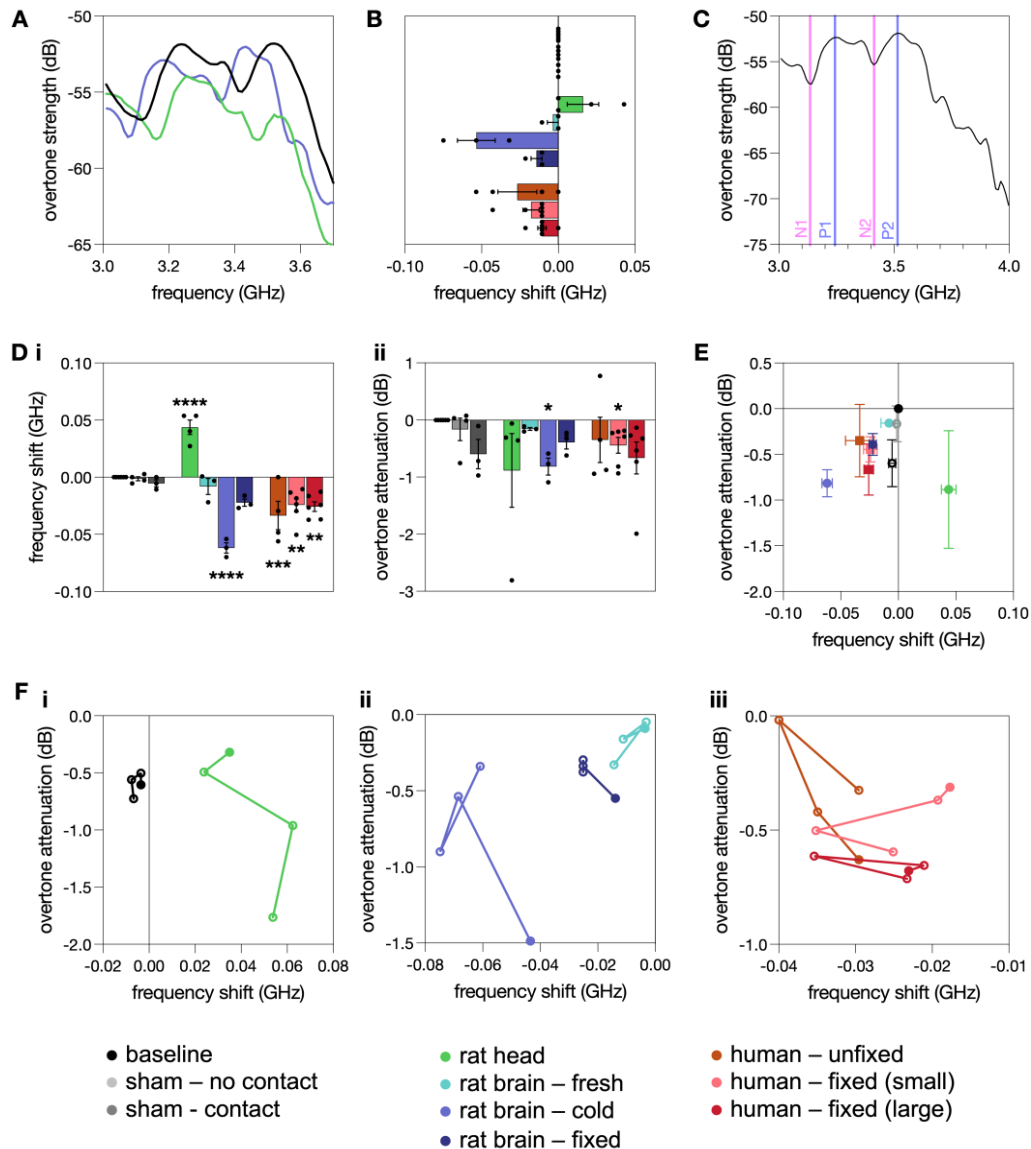

**Fig. 7. Impact of biological tissue on overtone strength. (A)** Representative overtone strength dynamics for baseline blank μBot (black trace), severed rat head (green trace), and explanted rat cold brain (purple trace) over the ME frequency range. **(B)** Overall overtone strength frequency shift identified by cross-correlation over the ME range between each condition and its respective baseline. **(C)** Identification of main overtone strength features, consisting of two negative (N1 and N2) and two positive (P1 and P2) peaks. **(D)** Quantification of frequency shift **(i)** and overtone attenuation **(ii)** of the identified signal features. Data represented as mean ± SEM, with individual samples scattered in black dots. *$p < 0.05$, **$p < 0.01$, ***$p < 0.001$, ****$p < 0.0001$ in relation to baseline using one-way ANOVA followed by Dunnet's multiple comparison test. **(E)** 2D representation of attenuation and frequency shifts caused by the different biological (and sham) samples. **(F)** Decomposition of sample impact on the different identified signal features: **(i)** sham control and severed rat head, **(ii)** explanted rat brain samples, and **(iii)** autoptic human samples. The filled dot corresponds to N1, sequentially connected to P1, N2, and P2.

### *Data communication and MRI Compatibility*

To evaluate the wireless capabilities of μBots in a realistic biomedical communication scenario, we established a transmission experiment at 3.25 GHz in which a 1.96 MB sonogram video with heartbeat audio was delivered in real time using a software-defined radio (SDR) platform and a pair of transmitting and receiving μBots, for the first time (Fig. 8A). Transmission followed the Digital Video Broadcasting-Satellite (DVB-S) standard with Quadrature Phase Shift Keying (QPSK) modulation at a symbol rate of 500 kilo symbols per second. A Root Raised Cosine filter with a roll-off factor of 0.35 yielded an occupied bandwidth of 675 kHz. Unlike conventional electrical antennas of comparable size, which are typically constrained to narrowband operation and require tuning to their environment, the μBots exhibited intrinsically wideband characteristics, readily accommodating this modulated signal without distortion or detuning. This wideband response ensured that antenna behavior did not limit the experiment, allowing performance to be determined entirely by the digital modulation and SDR hardware. The baseband waveform was generated at eight samples per symbol, corresponding to a digital sampling rate of 4 megasamples per second. This oversampling enabled accurate pulse shaping, stable synchronization, and reliable filtering while remaining within the real-time streaming limits of the B205i SDR platform. The local oscillator for both transmitter and receiver was fixed directly at 3.25 GHz, and although this direct-on-frequency approach introduced a small DC artifact at baseband, the robustness of DVB-S QPSK demodulation rendered the effect negligible. The average received power was obtained by integrating the measured spectrum over a bandwidth b centered at $f_0$ (via $\int_{f_0-\frac{b}{2}}^{f_0+\frac{b}{2}} P(\omega)d\omega$) under controlled in-plane and out-of-plane misalignments of the ME-RF link, and benchmarked against conventional RF-RF links. The RF antenna was a log microstrip PCB with the magnetic field oriented in-plane. Under

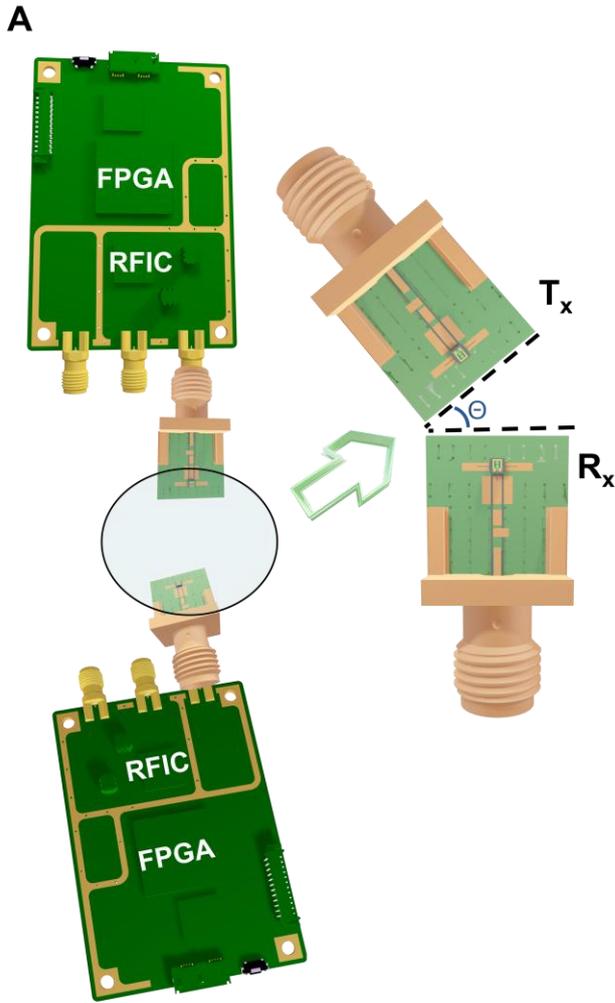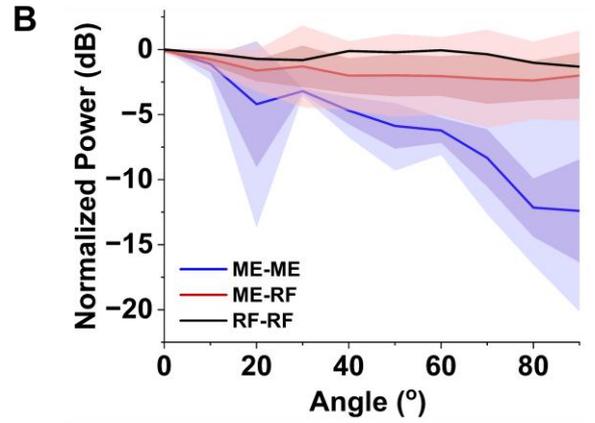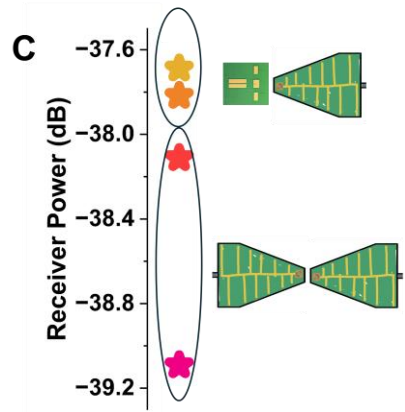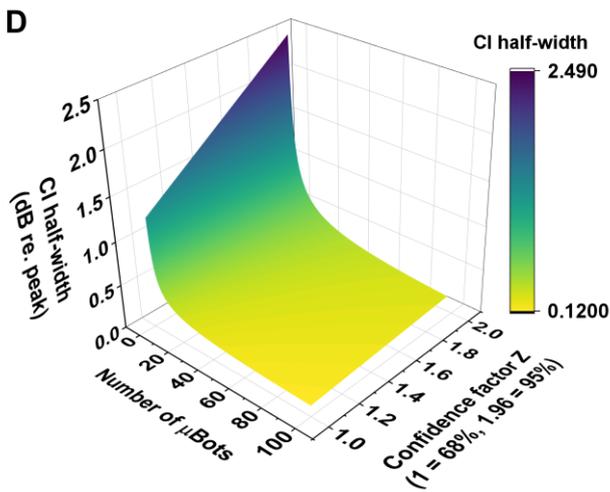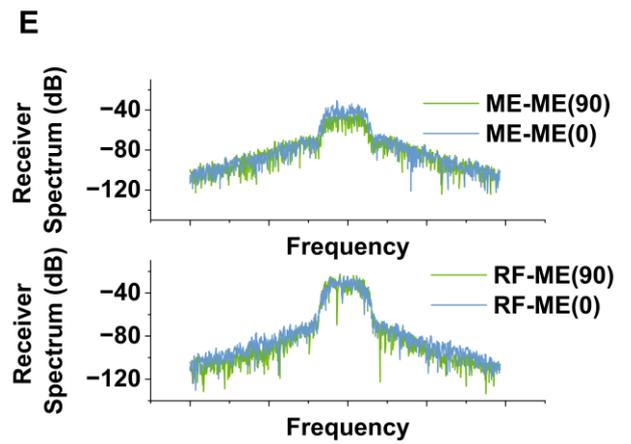

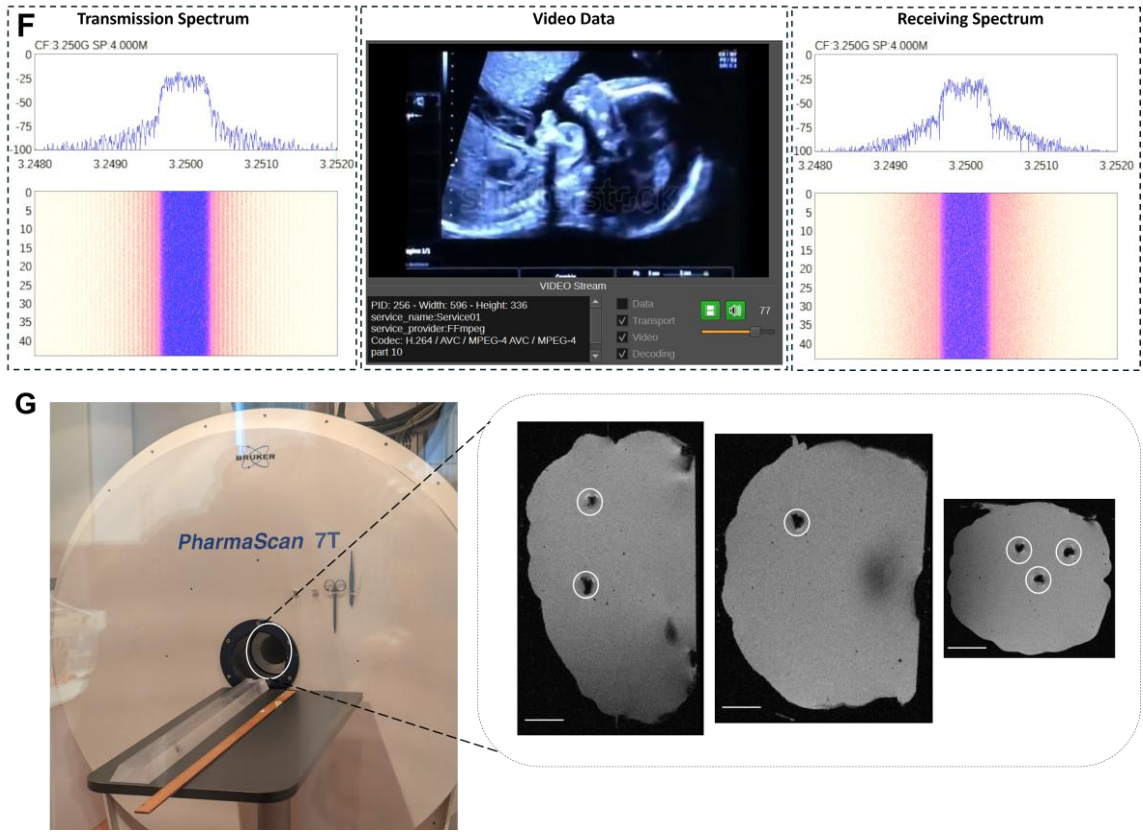

**Fig. 8.** Prototype link hardware demonstrating wireless audio-visual telemetry **(A)** Representational schematic of μBot (packaged OUWB-ME antenna) systems mounted between the transmitter and receiver front-ends (SI Note 1). **(B)** In-plane misalignment: normalized received power versus rotation angle for ME-ME (blue), ME-RF (red) and RF-RF (black) configurations. Shaded bands indicate confidence intervals (darker shade: 68%, lighter shade: 95%). **(C)** Out-of-plane tilt (0°, 90°): receiver-power distributions for ME-RF (top) and RF-RF (bottom) under identical tilts; ME-RF clusters higher and tighter, indicating reduced sensitivity to out-of-plane misalignment. **(D)** Misalignment heat-map indicating the number of μBots, N, required to obtain RF-like stability. CI half-width (dB re. peak) as a function of N and confidence multiplier, Z (1 = 68%, 1.96 = 95%) with the surface capturing $1/\sqrt{N}$ averaging. **(E)** Spectral fidelity of ME-ME (top) and ME-RF links (bottom) at (0°, 90°) overlap (center frequency: 3.25 GHz) showing minimal angle dependent-distortion of the DVB-S/QPSK waveform. Overlaid spectra at 0° and 90° confirm that hybrid links preserve the transmitted spectrum while benefitting from the misalignment robustness in **(C)**. **(F)** Real-time audio-visual sonogram streaming over a magnetoelectric link. The receiver is tuned to 3.25 GHz (span 4.0 MS/s); the left and right panels show the RF spectrum and waterfall for the transmitter and receiver, respectively. In both cases the waveform exhibits the same Root Raised Cosine shaped occupied band (0.675 MHz) centered at 3.25 GHz, with a stable vertical stripe in the waterfall indicating continuous lock. Center: decoded ultrasound sonogram with heart beat audio confirming end-to-end transport (SI Video 9, SI Note 2). **(G)** MRI of OUWB-ME embedded in Agar. Left: Bruker PharmaScan 7T system and phantom placement. Right: XY, XZ and YZ axial slices with device locations circles after 10-hour scan. The OUWB-ME antennas produce small, localized contrast "voids" at the implant sites without gross geometric distortion or large-scale signal non-uniformity, indicating practical MRI compatibility. Scale bar: 0.4 mm.

out-of-plane misalignment, a heterogeneous ME-RF pairing exhibited a markedly shallower power change than the RF-RF

baseline. In the measurement shown (top two markers, ME-RF; bottom two, RF-RF), the ME-RF points cluster higher and tighter on the power axis, indicating lower sensitivity to out-of-plane tilt than an RF-RF link under identical conditions (Fig. 8B). We ascribe this robustness to the magnetic-field-dominated near-field coupling provided by the ME element, which mitigates the polarization and orientation penalties that degrade purely RF pairs.

In terms of angular misalignment, the contrast between ME-RF and RF-RF can be understood by examining their confidence intervals and the number of ME elements required to achieve comparable robustness. The per-angle standard deviation was calculated directly from the measured response at each angular position, quantifying the spread of values around the mean for both ME-RF and RF-RF curves. A single ME element exhibits a per-angle standard deviation of approximately 1.27 dB, whereas an RF element (RF-RF) shows a lower spread of about 0.43 dB, reflecting its inherently greater stability. Yet, because ink-bonded ME elements occupy only ~1 mm² of printed circuit board (PCB) area compared to ~147 cm² for the RF antenna, a reduction of almost five orders of magnitude, they can be densely arrayed, and their effective standard deviation decreases as $1/\sqrt{N}$, where N is the number of combined elements. From this scaling, about nine ME elements are sufficient to reduce ME-RF's angular sensitivity to the same level as a single RF element. Importantly, the stability of nine ME elements is already close to that of an RF antenna, yet their combined footprint remains below 0.01 cm². More stringent requirements highlight this advantage further: to confine the 68% confidence interval within ± 0.5 dB requires 6 elements, achieving ± 0.2 dB requires 40 elements, and ± 0.1 dB requires 160 elements. At the 95% confidence level, the corresponding numbers are 25, 154, and 616 elements for ± 0.5, ± 0.2, and ± 0.1 dB, respectively. Extending the analysis to the ME-ME case with the widest variation, the measured mean and 95th-percentile per-angle standard deviations (1.78 dB and 4.44 dB, respectively) imply 16 elements on average and 100 elements for worst-angle robustness to match the RF-RF trend, still only 0.16 -1.0 cm² of ME footprint. These results demonstrate that although a single ME receiving element is more misalignment-sensitive than RF, the ability to integrate and average tens to hundreds of μBots within an area orders of magnitude smaller than that of a single RF antenna allows ME-RF not only to reach but to surpass RF-RF in robustness, while heterogeneous ME-RF pairings already display reduced out-of-plane sensitivity. Together, these findings establish a pathway for highly stable, miniaturized biomedical telemetry links where angular tolerance is critical.

Considering a federated application of μBots, MRI compatibility must be insured due to the presence of ferromagnetic materials, metallic layers and inks. Randomly dispersed clusters of OUWB-ME underwent 10-hour 7T MRI scans to gauge the movement of devices, if any. T1-weighted images exhibited only minor edge artefacts around the chip, insufficient to confound interpretation of *in vivo* data. Segmentation masks delineated chips with the expected size and morphology, without intrachip artefacts. Difference (ΔPos) images showed a net downward displacement between time points with no component along $B_0$, consistent with gravitational settling rather than magnetic forces. ΔPos encodes the voxel-wise mask difference (+1, present only at the first scan; 0, unchanged; -1, present only at the second) (SI Videos 5-8).

# Discussion

In this study, we report OUBW-ME antennas with bandwidth of 22.6 GHz, established through fabrication, characterization and validation. AlN is biocompatible with dissociated rat neurons *in vitro*, while FeGa induces cell death unless isolated by encapsulants such as Parylene C. Regarding the ex vivo µBot measurement, although preliminary, these experiments constitute, to our knowledge, the first systematic quantification of how biological tissue reshapes the transmission and reflection characteristics ME antennas. Across preparations, we observe reproducible spectral features that identify discrete frequency windows, most prominently near 3.3 and 3.9 GHz, likely to be preferential for reliable wireless links in tissue-embedded devices. These observations will directly inform the frequency bands implemented in forthcoming µBot technology. As neurotechnology increasingly pivots toward implantable, wirelessly operated systems, the need for rigorous, mechanistic characterization of biological-electronic interactions and interfaces becomes acute. Our ongoing measurements expand this emerging evidence base and will guide future device integration and the design of closed-loop communication strategies within the µBot federation. The data further reveal both conserved and divergent responses across species and tissue-preparation states, motivating additional work to resolve their biological origins. In this context, overtone frequency shifts are governed primarily by tissue-induced damping and by acoustic reflections to and from the device, whereas overtone amplitudes provide a quantitative readout of loss, decreasing as damping increases. Collectively, these results offer a framework for selecting operational bands and interpreting overtone structure in biologically constrained ME systems.

Single-element analyses often cast ME antennas as orientation-tolerant because they couple through the magnetic near field and avoid platform currents, but this intuition breaks down in federated arrays (*45*). Each ME element acts as a magnetic dipole whose axis is fixed by the DC-bias. Coupling is strong when this axis aligns with the receive field and collapses when orthogonal, so even small inter-element bias scatter accumulates as polarization and phase mismatch. In a federation, those mismatches are further amplified, eroding the native stability improvement expected from an individual antenna on-chip and unpackaged. Tissue further imposes nonuniform acoustic loading, shifting overtone frequency and strength element by element, so the federation experiences unequal complex gains rather than a uniform misalignment penalty. Our work provides the first insight into this behavior.

By combining compact form factor with wideband operation, magnetoelectric antennas overcome the limitations of conventional narrowband miniature antennas and provide a fundamentally different platform for wireless biomedical communication. The successful delivery of a sonogram video highlights their potential to enable next-generation wireless systems for minimally invasive diagnostics, implantable monitoring, and portable health technologies, where reliable and efficient transfer of imaging data is essential. Looking forward, the demonstrated link can be readily scaled to higher symbol rates, higher-order modulation formats, and multi-megabit data streams. For example, increasing the symbol rate to 2 mega symbols per second with QPSK would yield a throughput of 4 Mbps, sufficient to support full-motion ultrasound video or higher-fidelity diagnostic imaging.

The future of wireless telemetry with federations of micron-scale ME antennas depends on their wide 22.6-GHz usable band. With the available headroom, a federation of ME antennas can split the spectrum among nodes or take rapid turns sending short bursts, creating, when required, very wide effective signals that deliver few-millimeter ranging precision and tens-of-picoseconds timing, while still supporting multi-gigabit data streams even if only a small fraction of the band is used at any moment. The breadth also buys resilience and the network can slide its center frequency to avoid tissue- or hardware-induced attenuation, run several parallel streams for robustness, and keep power and SAR low by duty-cycling and beamforming rather than pushing a single element hard. The ME federation can choose between precision, rate, and robustness on the fly, making practical, body-embedded telemetry realistic in ways narrowband implantable antennas cannot.

These results point toward the broader applicability of magnetoelectric antennas in enabling wideband biomedical data transfer and future clinical telemetric applications. Furthermore, by shedding light on tissue- and condition-dependent effects not only on antenna transmission and reflection, but also on acoustic magnetoelectric overtone frequency shift and signal attenuation, this study inspects this previously unexplored domain. These unexploited interactions have significant implications, not just for understanding biological dielectric and acoustic behavior, but also for the next generation of implantable and wearable systems that must reliably navigate complex tissue environments.

# Materials and Methods

*Antenna fabrication and characterization*

*OUWB-ME antenna fabrication:* All OUWB-ME antennas were fabricated on undoped (100) double sided polished Silicon wafers of resistivity 10000-100000 $\Omega$cm. The substrates were cleaned in Piranha followed by deposition of the bottom electrode (Ti/Pt; 10/70 nm). Aluminum Nitride and FeGa were deposited via RF sputtering in $N_2$/Ar and Ar, respectively. The deposition parameters in terms of sputtering power, pressure, and flow rates were optimized towards c-plane AlN and (110) FeGa, deposited under a peak in-situ magnetic field of 10 kA/m, measured via a Hall sensor (TLV493D-A1B6; Infineon). Vias in AlN were etched using AZ400K developer and the top electrodes (Pt; 100 nm), patterned via lift-off in silicate developers. Prior to every deposition run, in-situ Argon etching was carried out for the removal of residual oxides and nitrides. Parallel permanent magnets provided DC in-plane domain-biasing during sputtering which was confirmed by on-chip two-port transmission measurements when normalized by the response of control piezoelectric resonators.

*Packaging of OUWB-ME antennas towards μBots:* Each antenna, initially protected with AZ4562 photoresist, was diced into 1 mm dies using a diamond saw (DISCO Corporation, Tokyo, Japan). The dies were mounted on ceramic Rogers PCBs with a non-conductive epoxy and the resist stripped in acetone followed by an IPA wash. Planar electrical interconnects were formed using a shear-thinning (non-Newtonian), glycol-based silver nanoparticle ink (CL85; XTPL, Poland) dispensed through a 20-μm-inner-diameter nozzle. To prevent oxidation of the Fe-based spintronic layer, the assemblies were heated at 180°C under continuous Ar flow.

*RF Characterization of OUWB-ME antennas:* On-chip reflection, transmission and impedance parameters were experimentally derived with a precision vector network analyzer (Keysight N5245B) following short-open-load-thru (SOLT) calibration. The frequency resolution was carefully chosen to spectrally resolve the overtones and the IF bandwidth and power for all measurements were held constant at 1 kHz and 0 dBm, respectively. The one-port impedance parameters were derived according to $Z_{11} = \frac{1+S_{11}}{1-S_{11}} Z_0$ with the characteristic impedance being 50 Ω. The derived impedance data at the overtone and higher order harmonic region at ~8 GHz was utilized to determine the RF properties of the active materials via the modified Sittig's model. The domain biasing in the magnetostrictive layer was exploited via the difference-model between the magnetoelectric antennas and control piezoelectric devices making sure the spatial position of the probe stays the same between measurement. All two-port transmission parameters are normalized by the one-port reflection parameter of the commercial horn antenna (PowerLOG 70180; Aaronia) to account for noise. The horn antenna was connected to port 1 and the OUWB-ME Antenna to port 2.

*Modified Sittig's transmission line model:* The input impedance of the overtone structure, $Z_{in}$, is given by Eq. 1. The effective input impedance is the parallel combination of the impedances of $A_1$ and $A_2$ (Fig. S4). In overtone designs, acoustic waves disperse into the underlying layers, including the substrate, generating a ladder of overtones with spacing $\Delta f \sim v/2t_s$, where $v$ is the longitudinal sound velocity in the substrate and $t_s$ its thickness. The phase delay in the piezoelectric layer is $\gamma = \omega t/V_{\text{piezo}}$, with analogous treatment applied for the remaining layers. Acoustic impedance $Z_{ai}$ follows Eq. 2, using thickness-weighted velocities and densities. In multilayer electrodes, properties are defined by relative layer fractions. A double-sided polished silicon wafer fundamentally alters the acoustic boundary wherein the smooth backside reflects rather than dissipates. This reflection disrupts assumptions of substrate loss, introduces additional phase terms, and reshapes standing wave conditions. Resonance shifts, impedance distortion, and Q degradation follow. The substrate is no longer passive, it becomes an active cavity and must be treated as such.

$$Z_{in} = \frac{1}{j\omega C_0}\left[1 - \frac{k_1^2}{\gamma}\frac{(z_b + z_2)\sin\gamma + j2(1-\cos\gamma)}{(z_b + z_2)\cos\gamma + j(1+z_b z_2)\sin\gamma}\right] = \frac{Z_1 Z_2}{Z_1 + Z_2} \quad (1)$$

$$Z_{ai} = A\rho V_{\text{piezo}} \quad (2)$$

$$V_{\text{net}} = \frac{t_1}{t_1 + t_2}V_{Pt} + \frac{t_2}{t_1 + t_2}V_{FeGa}$$

$$\rho_{\text{net}} = \frac{t_1}{t1 + t2}\rho_{Pt} + \frac{t_2}{t_1 + t_2}\rho_{FeGa}$$

$$Z_b = jZ_{\text{top\_electrode,net}} \times \tan\gamma_{\text{top}}$$

$$z_{\text{Low}} = j\frac{z_{Si}\cos\gamma_{\text{bottom\_electrode,net}}\sin\gamma_{Si} + z_{\text{bottom\_electrode,net}}\sin\gamma_{\text{bottom\_electrode,net}}\cos\gamma_{Si}}{\cos\gamma_{\text{bottom\_electrode,net}}\cos\gamma_{Si} - (z_{Si}/z_{\text{bottom\_electrode,net}})\sin\gamma_{\text{bottom\_electrode,net}}\sin\gamma_{Si}}$$

Where $z_b = Z_b/Z_{ai}$, $z_2 = Z_{\text{Low}}/Z_{ai}$, $\gamma_{\text{top}} = \omega d_{\text{top}}/V_{\text{top}}$, $\gamma_{\text{bottom}} = \omega d_{\text{bottom}}/V_{\text{bottom}}$, $\gamma_{Si} = \omega d_{Si}/V_{Si}$, $Z_{\text{top}} = A\rho_{\text{top\_electrode}}V_{\text{top\_electrode}}$, $Z_{\text{bottom\_electrode}} = A\rho_{\text{bottom\_electrode}}V_{\text{bottom\_electrode}}$,

and $Z_{Si} = A\rho_{Si}V_{Si}$. It must be noted that the density and velocity of the top electrode includes the contribution from spintronic layer.

*Vibrating Sample Magnetometry.* Magnetic hysteresis loops of the samples were obtained using a Vibrating Sample Magnetometer (MicroSence VSM-EV9). A maximum magnetic field of +/- 5 kOe was applied in the plane of the film at 0º and 90º concerning the short axis of the sample. Measurements were also performed out-of-plane of the film with a maximum applied field of +/-20 kOe, to allow for full magnetization saturation.

*X-ray Diffraction.* The crystallography was via Empyrean, Malvern Panalytical, with $K_{\alpha 1}$ and $K_{\alpha 2}$ wavelengths at 1.540598 Å and 1.544426 Å, respectively, and of intensity ratio 0.5. The pulse height detection levels by 25% in Empyrean for accurate detection of Iron containing compounds that fluoresce with Cu radiation. The raw data was subsequently processed by carrying out background subtraction via the denoising methodology of Sonnerveld and Visser.

*X-ray Photoelectron Spectroscopy.* The surface properties of the sputtered films with the inclusion of the Al capping layer were analyzed via X-ray photoelectron spectroscopy (AXIS Supra+, Kratos Analytical) consisting of an Al Kα anode producing X-ray beams of energy 1.486 keV. The beam power is held at 450W leading to a penetration depth of ~ 9 nm. The deconvoluted fit of all spectra from the raw data is carried out via Voigt fitting of the Gaussian and Lorentzian functions.

*In vitro biocompatibility*

*Materials.* Poly(ethyleneimine) (PEI), Hank's balanced salts ($Ca^{2+}/Mg^{2+}$ free) (HBSS), kynurenic acid, trypsin, deoxyribonuclease (DNase), bovine serum albumin (BSA), $NaHCO_3$, $MgSO_4$, HEPES, NaCl, KCl, $Na_2HPO_4$, heat-inactivated horse serum, D-glucose, paraformaldehyde (PFA), Triton X-100, rabbit anti-β3-tubulin primary antibody, mouse anti-GFAP primary antibody, and DAPI (0.5 mg/mL) were purchased from Merck KGaA (Darmstadt, Germany). Cell counting kit-8 (CCK-8) was acquired from Dojindo EU GmbH (Munich, Germany). Ethanol (70%), minimum essential medium (MEM), gentamycin (50 mg/mL), GlutaMAX, donkey anti-rabbit 488 secondary antibody, donkey anti-mouse 594 secondary antibody, and Oregon Green BAPTA-1 were obtained from ThermoFisher Scientific (Waltham, MA, USA). Fetal bovine serum (FBS) was purchased from Euroclone (Milan, Italy). D(-)-2-amino-5-phosphopentanoic acid (D-AP5) was acquired from Hello Bio (Dunshaughlin, Ireland). For cell culture experiments, two different solutions were used for the dissection protocol: dissection solution, consisting of 9.52 g $L^{-1}$ HBSS, 4.2 mM $NaHCO_3$, 33 mM D-glucose, 200 μM Kynurenic acid, 25 μM D-AP5, 0.025% (v/v) gentamycin, 0.03% (w/v) BSA, 12 mM $MgSO_4$, and 12 mM HEPES; and digestion solution, composed of 137 mM NaCl, 5 mM KCl, 7 mM $Na_2HPO_4$, 25 mM HEPES, 200 μM Kynurenic acid, 25 μM D-AP5, and 4.2 mM $NaHCO_3$. Similarly, two different cell media were used: H-MEM, consisting of MEM, supplemented with 5% (v/v) heat-inactivated horse serum, 20 mM glucose, 0.1% (v/v) gentamycin, and 100 μM GlutaMAX; and N-MEM, with the same composition as H-MEM, except with 50 μM GlutaMAX.

*Primary rat neuronal cultures.* Before cell plating, all substrates were sterilized in 70% ethanol for 30 minutes, washed 3x with sterile ultrapure deionized water, and allowed to air dry. To enhance neuronal adherence, substrates were coated with 0.1% (w/v) polyethyleneimine (PEI) overnight at 37 ºC, followed by 3x washes with sterile ultrapure deionized water and air drying. Primary neuronal cultures were prepared from the cortical neurons of P0-1 Wistar rats. Animals were sacrificed by decapitation, and their brains extracted and placed in 5 mL of dissection solution. Following the removal of the meninges, the cortical area was isolated and separated into small pieces in 5 mL of fresh dissection solution. The tissue was then moved under a laminar flow cabinet, washed with digestion solution, and exposed to 0.5% (w/v) trypsin and 0.075% (w/v) DNase in digestion solution at 37ºC for 5 minutes. The reaction was then stopped by incubating the tissue with 0.1% (w/v) trypsin inhibitor in dissection solution for 10 minutes at 4ºC. Finally, the tissue was washed in dissection solution, placed in a solution of 0.06% (w/v) DNase in dissection solution, and mechanically dissociated. Cells were then centrifuged (100g, 5 minutes), after which the supernatant was discarded, and the pellet was resuspended in H-MEM. Following cell counting, the final cell suspension was plated on the prepared substrates or glass coverslips as the control condition, at a final density of 1500 cells mm$^{-2}$ in H-MEM, kept at 37ºC in a humidified environment of 95% air-5% $CO_2$. Every two days, half the volume of cell medium was replaced with fresh media. Importantly, after 7 DIV, N-MEM was used. All experimental procedures were approved by local veterinary authorities and performed in accordance with Italian law (decree 116/96) and EU guidelines (86/609/CE, 2007/526/CE and 2010/63/UE). Animal use was approved by the Italian Ministry of Health. Efforts were made to minimize suffering and the number of animals used.

*Cell viability assay.* At different time points (7, 14, 21, and 28 DIV), cell medium was renewed and supplemented with 10% (v/v) of CCK-8. After 2 hours of incubation, 100 μL of cell medium was collected and its absorbance was read at 450 nm using a microplate reader (Multiskan FC, Thermo Scientific).

*Immunostaining.* At 14 DIV, cells were fixed with 4% (w/v) PFA in PBS for 30 minutes and washed 3x with PBS. Samples were permeabilized and blocked using 0.1% (v/v) Triton X-100 and 5% (w/v) FBS in PBS for 1 hour, followed by overnight incubation at 4ºC with primary antibodies (1:300). Then, samples were incubated with the respective secondary antibodies (1:500) for 2 hours and exposed to DAPI (1:500) for 30 minutes, both at room temperature and protected from light. Finally, samples were mounted on glass slides and imaged using an inverted fluorescence microscope with a 20x objective.

*Calcium imaging.* At 21 DIV, cell medium was removed and replaced with 5 μM Oregon Green BAPTA-1 in N-MEM. Following a 2-hour incubation at 37ºC, samples were transferred to fresh medium and imaged with an inverted fluorescence microscope equipped with a 10x objective. Each sample was imaged at 5 different fields of view, with each recording lasting 180 s with a frame rate of 5 Hz.

*Data analysis.* For the cell viability assay, every independent experiment was performed with technical duplicates and at least three independent experiments were performed for each condition. Given the different surface area of the glass coverslips and the silicon wafers, the absorbance measurements were normalized to substrate surface area. For immunostaining experiments, five random fields were acquired for every condition in every independent biological replicate. Immunostaining quantification was performed using custom MATLAB scripts, where the percentage of cell coverage was determined by the ratio of positive pixels in each channel following image binarization. The cell density was calculated based on the number of nuclei found in each analyzed field. For calcium imaging experiments, the acquired videos were also analyzed using custom MATLAB scripts, where all individual cells were identified, and their mean fluorescence signal was used as the calcium transient trace.

*Ex vivo transmission and reflection measurements*

*Experimental setup.* The μBot was mounted on a PCB and placed at a fixed distance of 16 cm from a commercially available horn antenna (PowerLOG 70180; Aaronia), to be within the radiating near field region of the horn antenna. The μBot was placed in a custom isolation box interference with microwave absorbers to minimize external. The horn antenna was oriented in such a manner that it exhibited minimal electrical coupling to ground thus reducing parasitic effects while maintaining a high in-plane H-field. Transmission ($S_{12}$) and reflection ($S_{22}$) parameters were measured across the magnetoelectric overtone operational range (3-4 GHz), using six different μBots. VNA calibration was carried in the entire testing frequency range of 0.7-5 GHz, with a port power of 0 dBm and an IF bandwidth of 100 Hz. The horn antenna was connected to port 1 and the μBot integrated with the *ex vivo* samples to port 2.

*Animal sample preparation.* Postnatal day 28 (P28) Wistar rats were anesthetized and decapitated, after which the entire head was placed on the ME antenna with a thin layer of Parafilm for separation. Transmission and reflection measurements were then acquired. Subsequent measurements were performed after sequentially extracting the brain, immersing it in ice-cold artificial cerebrospinal fluid (aCSF) for 10 minutes, and finally fixing it in 4% paraformaldehyde (PFA) in PBS for 24 hours. Four animals were used in total. All experimental procedures were approved by local veterinary authorities and performed in accordance with Italian law (decree 116/96) and EU guidelines (86/609/CE, 2007/526/CE and 2010/63/UE). Animal use was approved by the Italian Ministry of Health. Efforts were made to minimize suffering and the number of animals used.

*Human sample preparation.* Access to human neocortical autoptic samples was obtained, upon approval by the Local Ethical Committee of the Univ. of Modena & Reggio Emilia, through the Forensic Medicine Institute of the same university. One sample was formalin-fixed post-autopsy and subdivided into a "large" and a "small" cortical piece. The second was harvested during autopsy and frozen immediately, only thawed prior to the experiment. Measurements were performed on both fixed and unfixed tissues.

*Controls.* As sham controls, an inert modeling clay of equivalent weight to the rat head was used. Both contact and non-contact configurations were tested to disambiguate effects due to weight or dielectric properties.

*Signal processing and analysis.* Reflection ($S_{22}$) and transmission ($S_{12}$) magnitudes were calculated as the absolute value of the complex (real + imaginary) signal components. $S_{12}$ was normalized against the reflection magnitude ($S_{11}$) of the horn antenna to rule out any external fluctuations associated with the horn antenna. The magnitude of each tested sample was converted in fold-change in relation to the respective control, to minimize baseline differences between the different tested μBots. Differences in transmission and reflection were first identified by running a series of frequency-by-frequency t-tests between each condition and its control, with p-values adjusted by the False Discovery Rate (FDR) approach (Benjamini-Krieger-Yekutieli). Frequencies with consistent effects across samples were further analyzed using Kruskal-Wallis tests with FDR. To quantify reflection overtone strength throughout the tested frequency range, $S_{22}$ traces were denoised by subtracting the low frequency component of the trace to remove the overall signal dynamics and isolate the overtones. Overtone strength was then calculated as the peak-to-peak amplitude of the denoised signal. Overtone strength cross-correlation between each sample and its respective control was used to determine the overall frequency shift associated with each sample. Furthermore, from the overtone strength dynamics, the two major peaks and troughs were identified and separately analyzed to identify frequency and feature specific differences, using one-way ANOVA followed by Dunnet's multiple comparison test.

*Wireless telemetry of sonogram with heartbeat audio:* To quantify how orientation affects data transfer, we measured link gain while rotating the μBot transmitter-receiver pair in azimuth (in-plane) and elevation (out-of-plane) connected to two software defined radio transceivers (SRP B205mini-i; Ettus Research). The antennas were mounted with their planes parallel at a fixed 10 mm separation and the transmitter normal defined 0°. For the azimuthal sweep, the receiver was rotated about its surface normal from 0° to 90° in 10° increments while transmit power and separation were held constant, and the received power was recorded in SDRangel. Continuous spectra and video were monitored to verify lock. . The transmitter employed a low-pass filter (LPF) cutoff of 0.8-1.0 MHz, slightly exceeding the occupied bandwidth to preserve signal fidelity while suppressing spectral leakage. The receiver bandwidth was configured more broadly, at 1.0-1.2 MHz, ensuring complete capture of the transmitted spectrum, including the filter roll-off and any minor frequency offsets, while still excluding excess noise. Transmit gain was set to 35-40 dB to avoid amplifier saturation and receive gain to 55-65 dB to maximize the signal-to-noise ratio.

*MRI Compatibility test:* Six OUWB-ME antennas were placed into 2% agar (MSK, UK) before it had reached its gelling temperature and left to set. Magnetic resonance images were acquired on a 7T PharmaScan scanner with a 1P T8102 1H volume coil (Bruker Biospin, Germany). 3D T1 weighted (T1w) fast low angle shot (FLASH) images were acquired, with two shorter scans at a lower resolution (TR/TE = 20/5 ms, FA = 10º, FOV = 30 × 21 × 21 mm, imaging matrix = 300 × 210 × 210, Av = 2) acquired at an 10 hour, 8 minutes interval, and a high resolution scan (TR/TE = 50/4.8 ms, FA = 10º, FOV = 26 × 17 × 20 mm, imaging matrix = 350 × 230 × 270 mm, Av = 10). Images were processed in MATLAB (MathWorks, USA), with masks created by defining voxels that contained less than 10% of the maximal signal value for the entire image. Delta images were calculated by the difference of the initial image or mask minus the final image.

**Bio-integrated µBots with Overtone Ultra-Wideband Magnetoelectric Antennas for Wireless Telemetry**

Mahdieh Shojaei Baghini[1], Adam Armada-Moreira[2], Alessio Di Clemente[2,3], Dibyajyoti Mukherjee[1], Afesomeh Ofiare[4], Jonathon Harwell[5], Mary Dysko[6], Luana Benetti[8], Declan Bolster[9], Laura Mazon Maldonado[1], Dayhim Nekoeian[1], Moreno Maini[7], Mostafa Elsayed[1], Rossana Cecchi[2], Ricardo Ferreira[8], Jeff Kettle[5], Sandy Cochran[6], William Holmes[9], Carlos Garcia Nunez[1], Luca Selmi[7], Nicola Toschi[10,11], Michele Giugliano[2,3,12], Hadi Heidari[1]

[1] Microelectronics Laboratory, James Watt School of Engineering, University of Glasgow, G128QQ, United Kingdom

[2] Department of Biomedical, Metabolic and Neural Sciences, University of Modena and Reggio Emilia, Via Campi 287, 41125 Modena, Italy

[3] International School for Advanced Studies (SISSA), Via Bonomea 265, 34136 Trieste, Italy

[4] Centre for Advanced Electronics, University of Glasgow, G128QQ, United Kingdom

[5] James Watt School of Engineering, University of Glasgow, G128QQ, United Kingdom

[6] Centre for Medical & Industrial Ultrasonics, James Watt School of Engineering, University of Glasgow, Glasgow G12 8QQ, United Kingdom

[7] Department of Engineering "Enzo Ferrari", via Pietro Vivarelli 10, University of Modena and Reggio Emilia, 41125 Modena, Italy

[8] International Iberian Nanotechnology Laboratory (INL), Braga, Portugal

[9] School of Psychology and Neuroscience, College of Medicine, Veterinary and Life Science, University of Glasgow, Glasgow G61 1QH, United Kingdom

[10] Department of Biomedicine and Prevention, University of Rome, "Tor Vergata", Rome, Italy

[11] Department of Radiology, Athinoula A. Martinos Center for Biomedical Imaging, Massachusetts General Hospital, Harvard Medical School, Charlestown, MA, United States

[12] National Interuniversity Consortium of Materials Science and Technology (INSTM), Florence, Italy

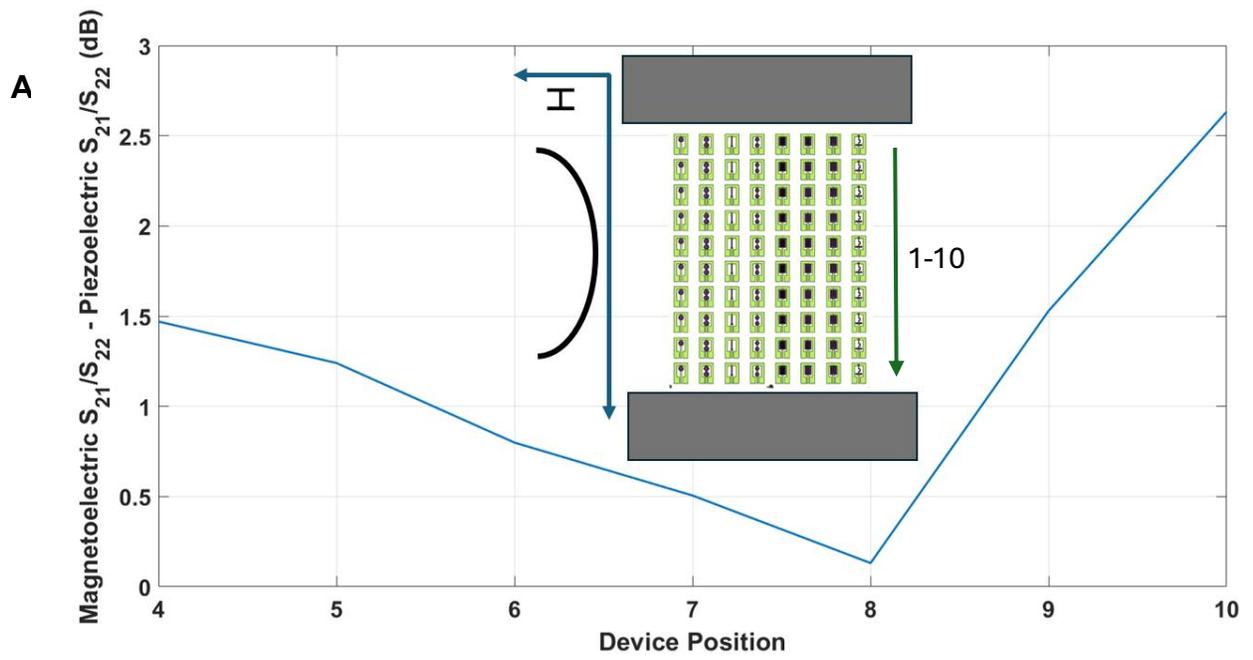

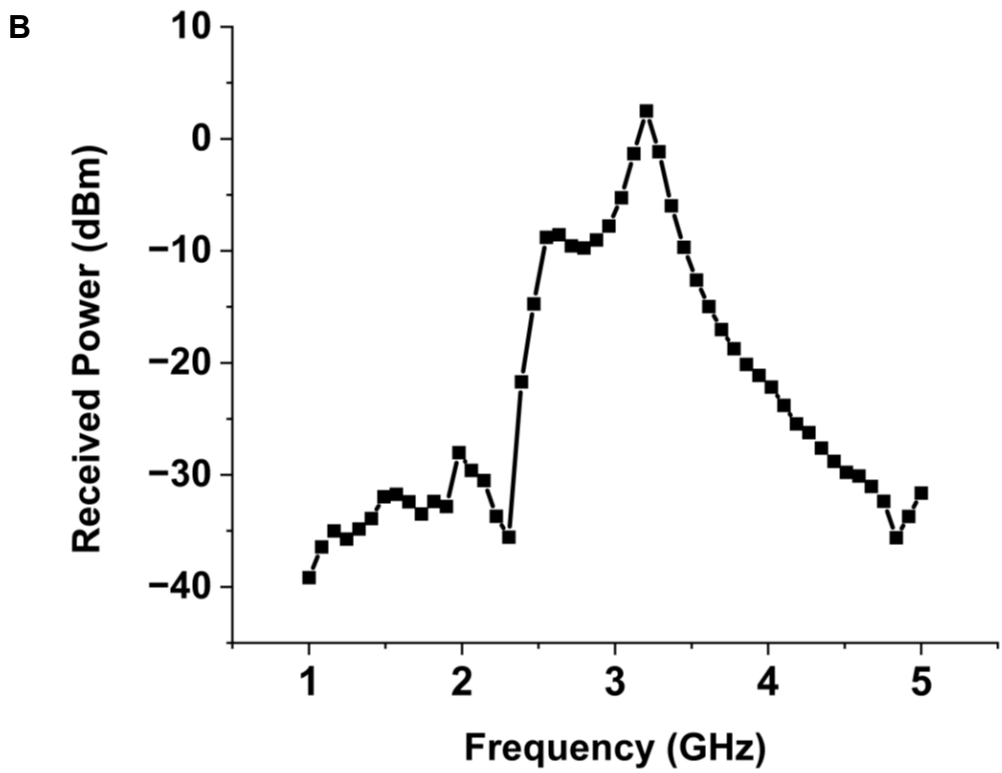

**Fig. S1A.** Variation in effective transmission of OUWB-ME antennas during overtone with respect to device position (shown by green arrow) and corresponding strength of magnetic field applied during deposition. The difference reaches near 0 at the center of the sample where the field is minimum. The variation is subtracted from control piezoelectric device response and normalized by one-port reflection parameter of the horn antenna. **Fig. S1B.** Raw received power peaking at ~3.2 GHz within the overtone range when the OUWB-ME antenna is connected as a receiver to a spectrum analyzer. The transmitting antenna is a horn antenna why the magnetic field oriented parallel to the easy-axis/in-situ DC field axis.

**Table S1.** RF mechanical properties of the materials composing the AlN/FeGa magnetoelectric antennas required for the modified Sittig's model. '-' indicates the properties that are not required in the first order model for determining the overtone frequency and coupling coefficients. The elastic and piezoelectric constants are in the longitudinal mode.

|  | Longitudinal acoustic velocity (m/s) | Density (kg/m³) | Dielectric constant | Elastic constant (GPa) | Piezoelectric constant (C/m²) |
|---|---|---|---|---|---|
| Silicon | 8430 | 2329 | - | - | - |
| Titanium | 6000 | 4400 | - | - | - |
| FeGa | 2828 | 7500 | - | 59.95 | - |
| AlN | 11000 | 3200 | 9 | 289 | 0.9 |
| Platinum | 3200 | 20000 | - | - | - |

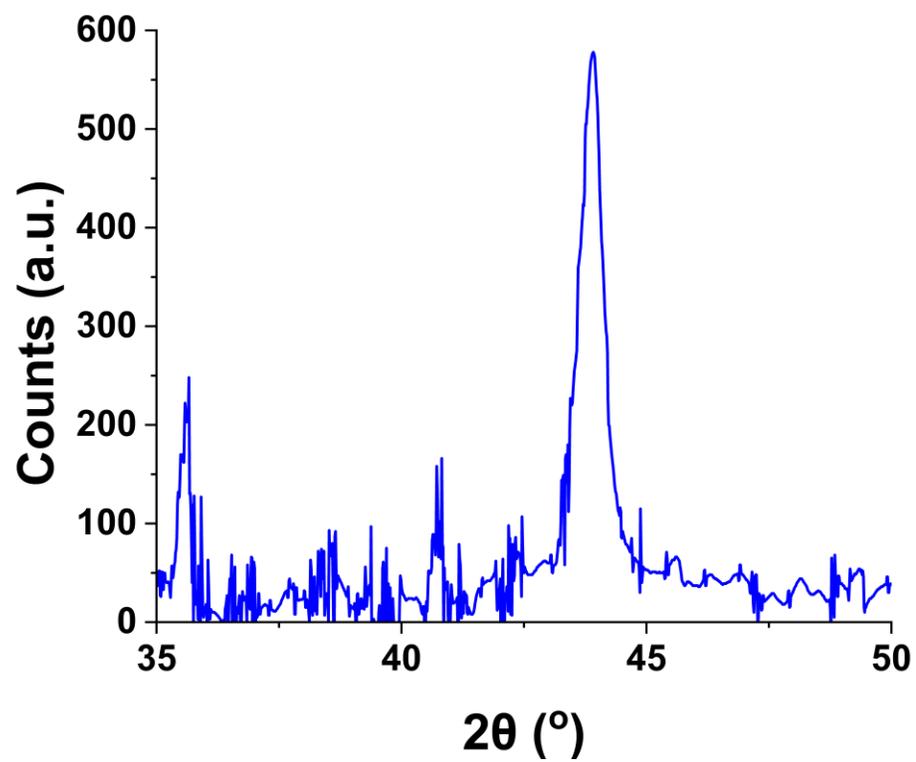

**Fig. S2**. XRD of sputtered FeGa (110) at 43.94º affiliated to increased magnetostriction due to co-existence of D03 and A2 phases.

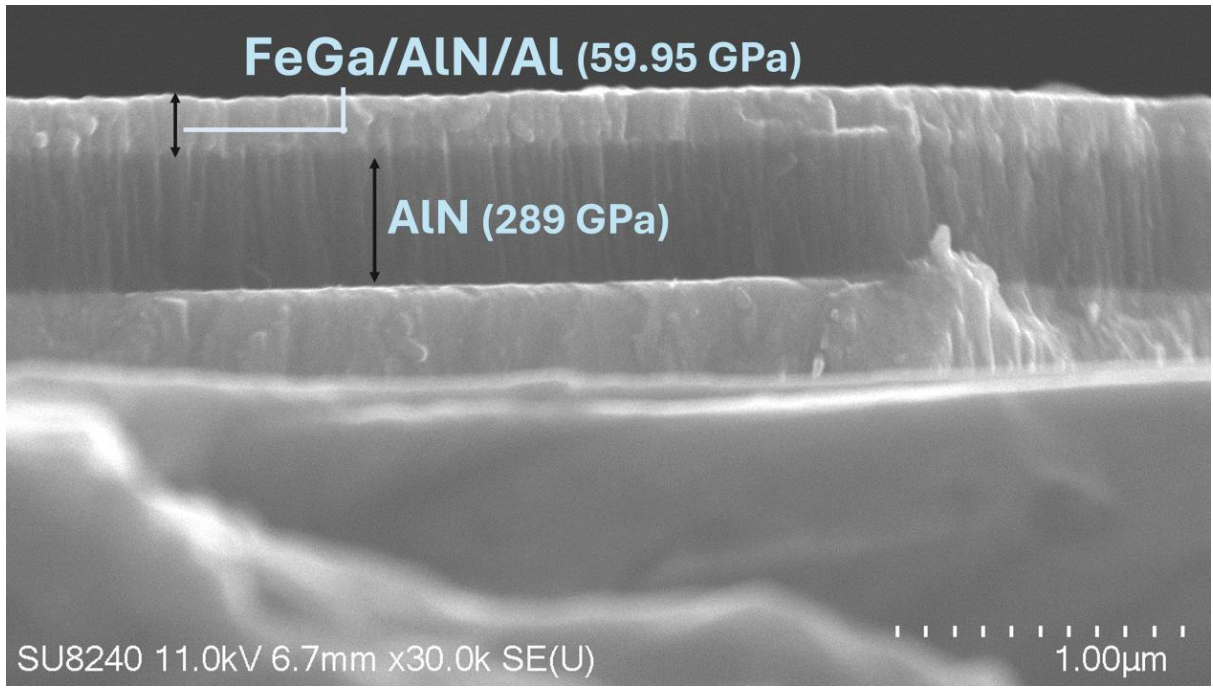

**Fig. S3**. SEM of ME films with columnar grains deposited on low-stress polished Si.

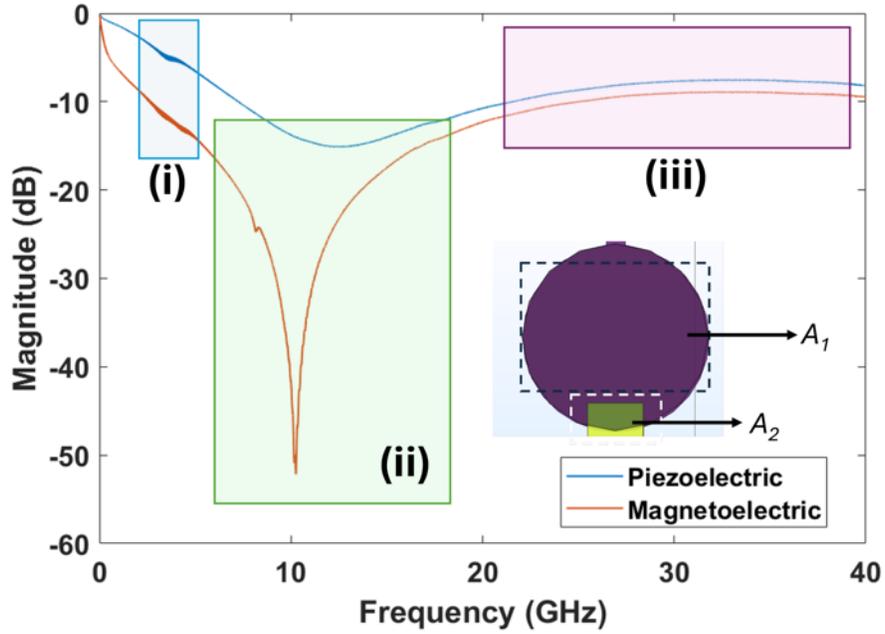

**Fig. S4.** Reflection coefficient ($S_{11}$) of piezoelectric resonators and OUWB-ME consisting of (i) high overtone region, (ii) spurious region and self-resonance, (iii) self-resonance. Inset highlighting the two distinct heterostructure stacks required for accurate modeling for derivation of RF electromechanical properties and stiffness.

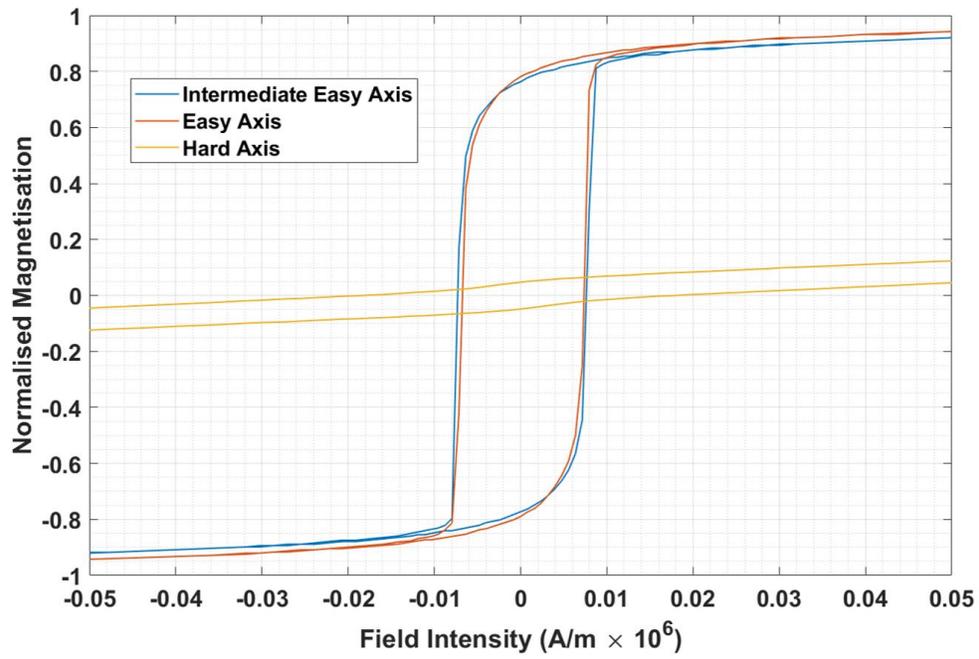

**Fig. S5**: In-plane and out-of-plane magnetic hysteresis loop of $Fe_{0.79}Ga_{0.21}$ film.

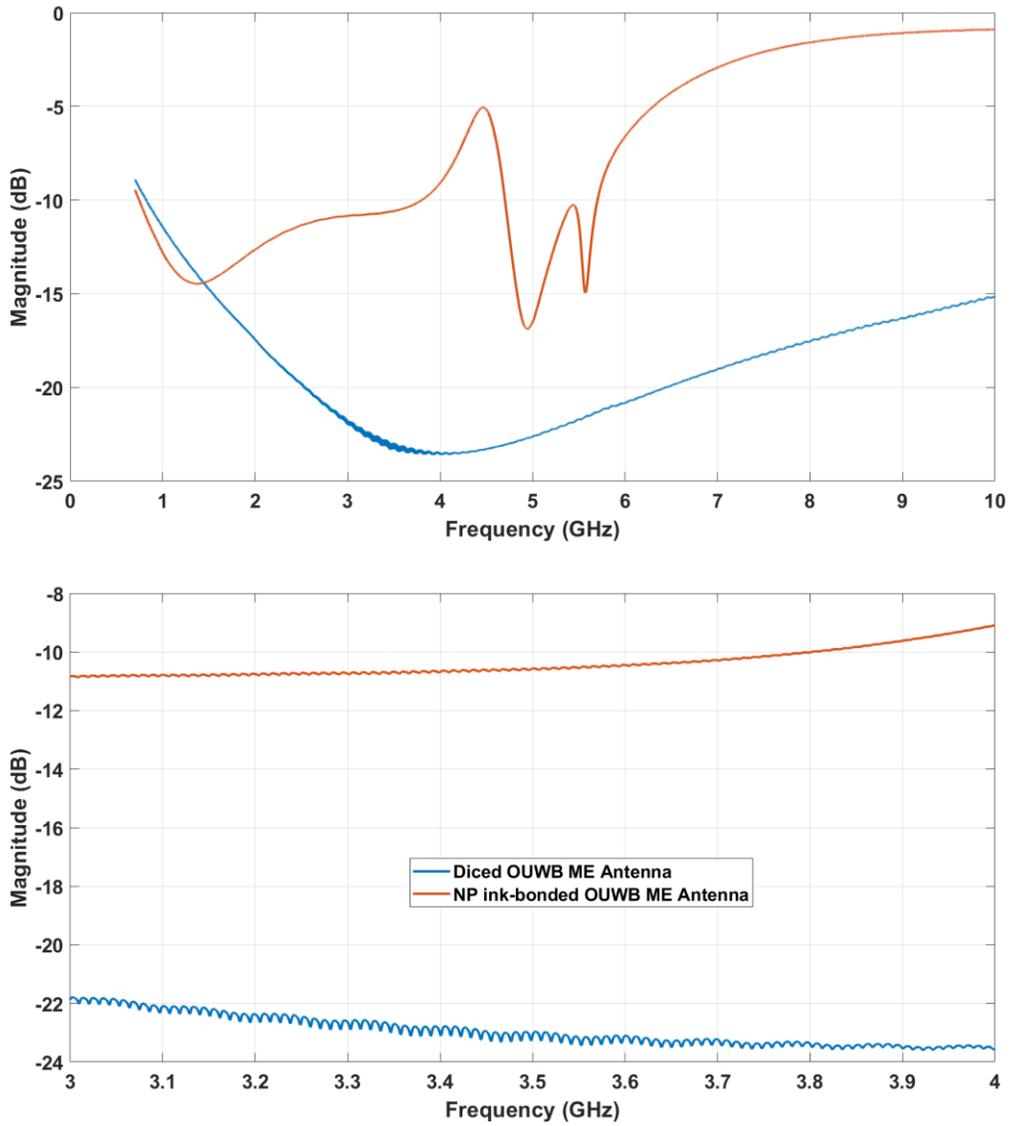

**Fig. S6**: Comparison of as-diced OUWB-ME antenna and ink-bonded OUWB-ME antennas (µBot). The phase shift due to the conductive ink and PCB traces introduces additional resonance peaks outside the magnetoelectric overtone range of interest spanning 3–4 GHz (more primarily 3.1–3.6 GHz).

**Table S2. Biological samples used in the µBot transmission and reflection measurements.**

| Sample | Description | Average weight (g) | N |
|---|---|---|---|
| Sham – no contact | Synthetic mass phantom, not in contact with antenna | 12.5 | 4 |
| Sham - contact | Synthetic mass phantom, direct contact | 12.5 | 3 |
| Rat head | Intact decapitated head | 12.5 | 4 |
| Rat brain – fresh | Immediately following extraction | 1.71 | 4 |
| Rat brain – cold | 10 min in ice-cold aCSF | 1.7 | 3 |
| Rat brain – fixed | 24h in PFA | 1.89 | 4 |
| Human – fixed (small) | Formalin-fixed small fragment | 1.9 | 6 |
| Human – fixed (large) | Formalin-fixed large fragment | 11.38 | 6 |
| Human – unfixed | Frozen post-autopsy and thawed immediately prior to measurement | 4.72 | 4 |

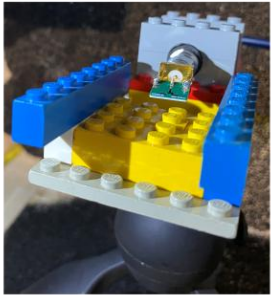

Non-conducting mount

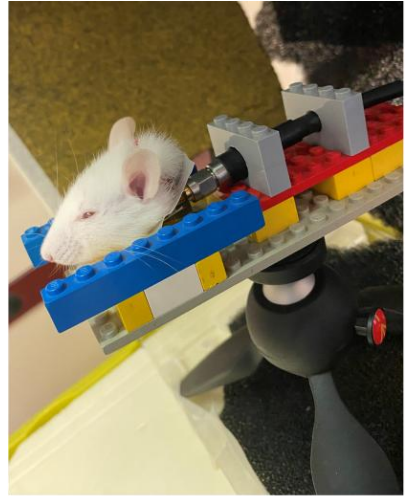

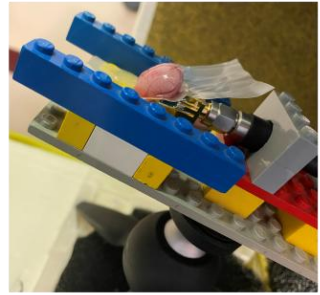

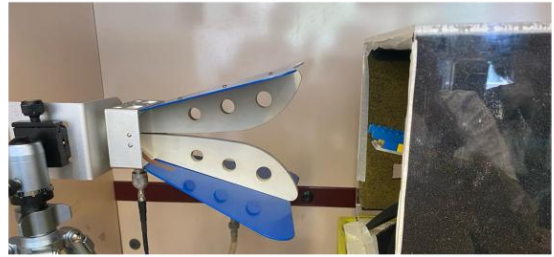

Horn antenna with H-field parallel to ground

**Fig. S7**: Experimental setup of *ex vivo* reflection and transmission measurements of µBots.

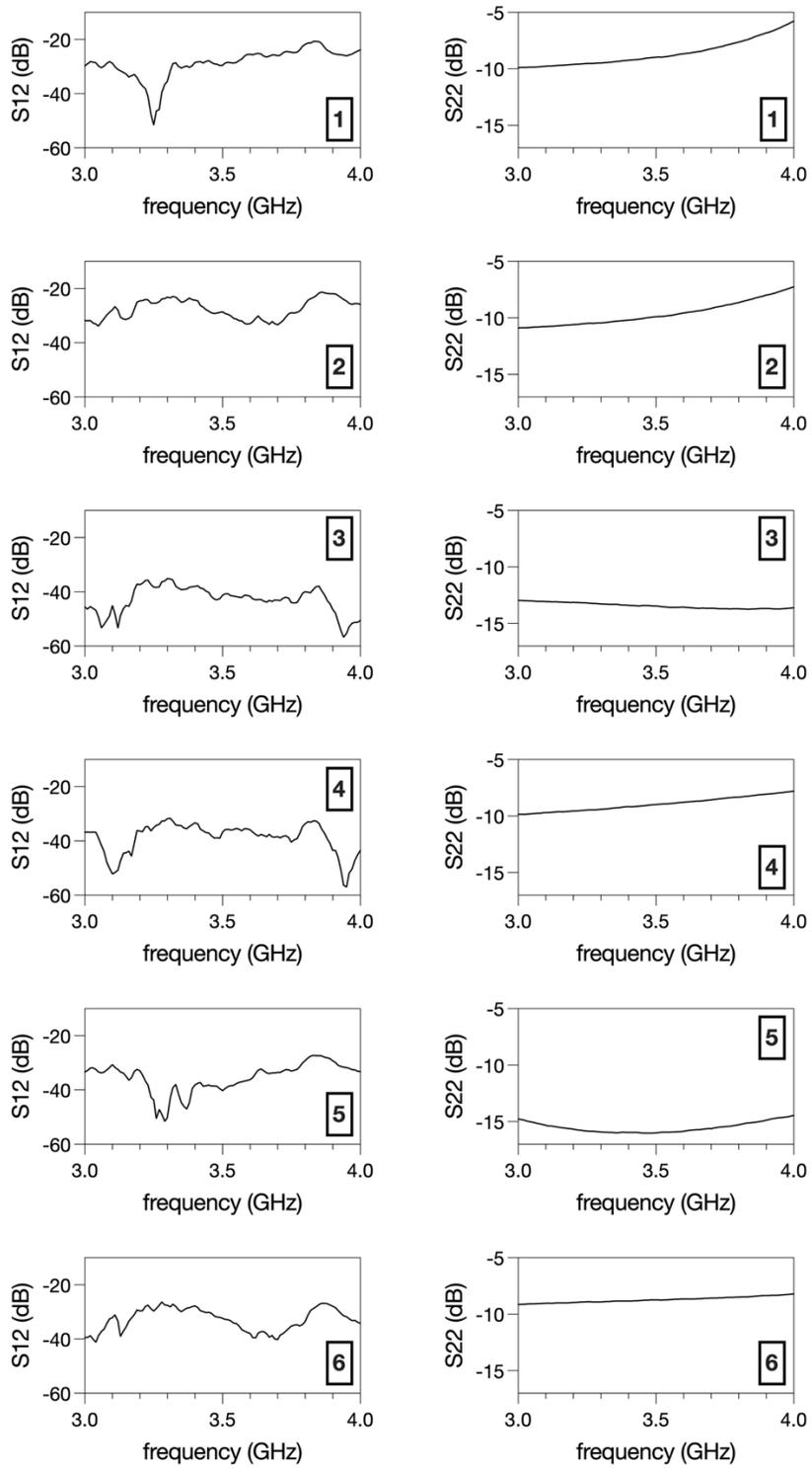

**Fig. S8.** Blank $S_{12}$ and $S_{22}$ measurements for all μBots.

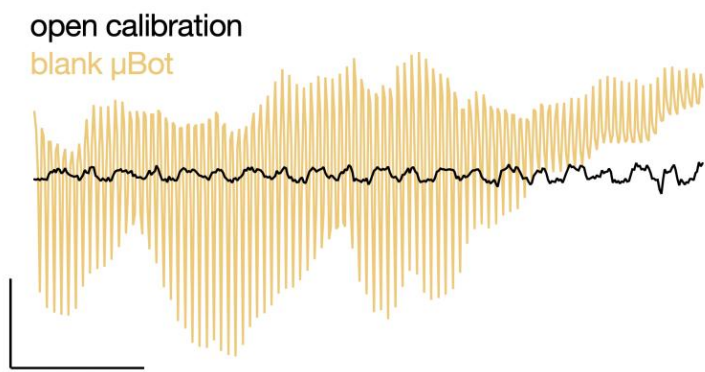

**Fig. S9.** Isolated overtones (fast oscillation component) for open calibration and blank μBot measurements. X-scale bar = 0.2 GHz. Y-scale bar = 0.001 absolute magnitude.

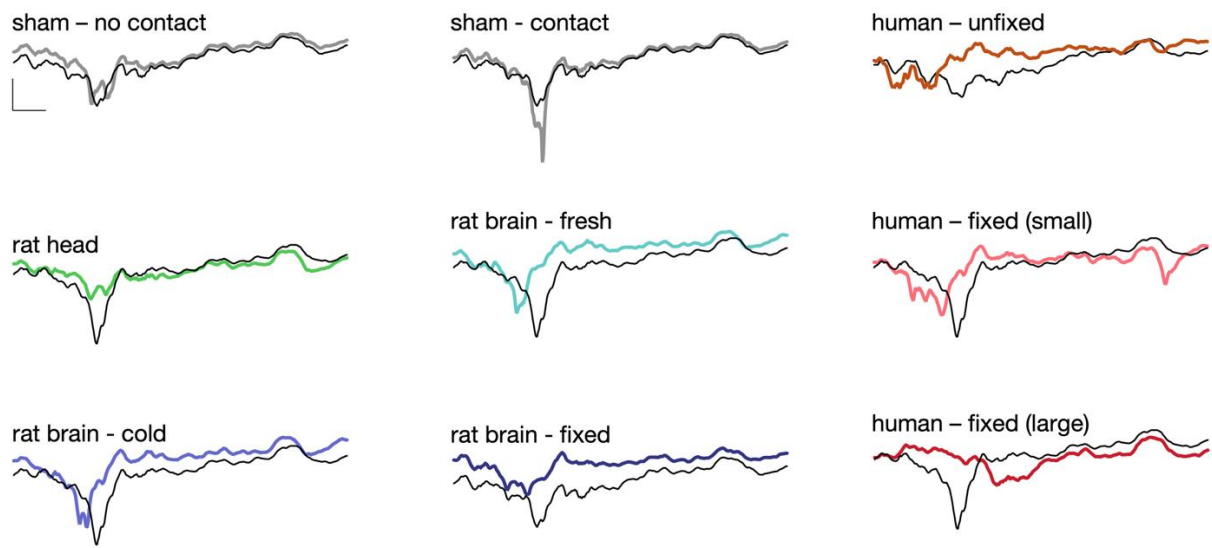

**Fig. S10. Representative $S_{12}$ traces for all tested conditions.** Respective blank measurement in black. X-scale = 0.1 GHz. Y-scale bar = 10 dB.

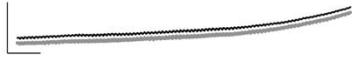
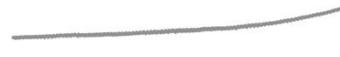
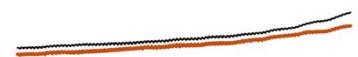
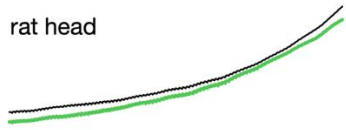
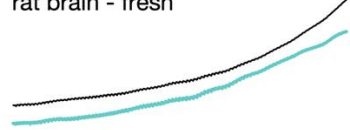
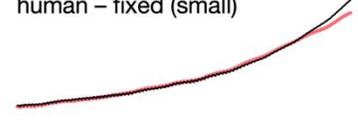
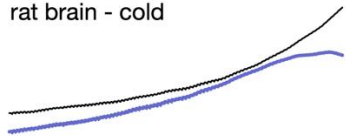
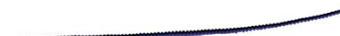
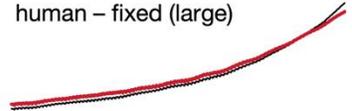

**Fig. S11. Representative $S_{22}$ traces for all tested conditions.** Respective blank measurement in black. X-scale = 0.1 GHz. Y-scale bar = 2 dB.

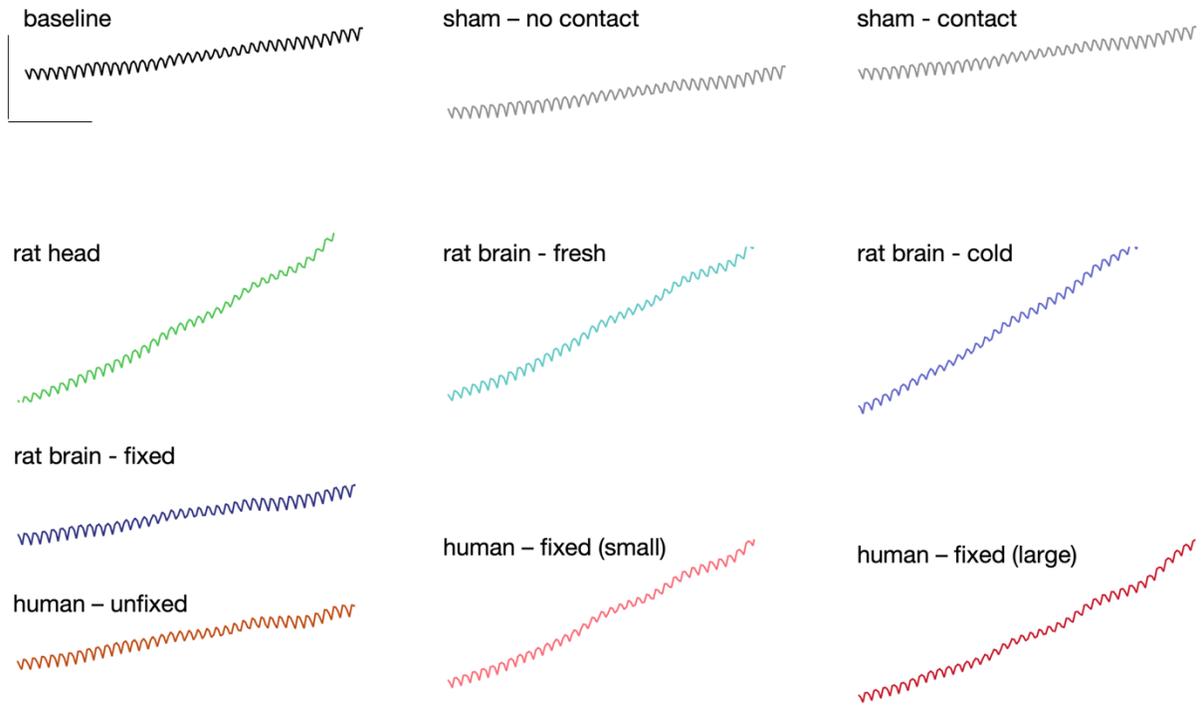

**Fig. S12. Detailed representative S$_{22}$ traces for all tested conditions, highlighting overtones.** Respective blank measurement in black. X-scale = 0.1 GHz. Y-scale bar = 0.5 dB.

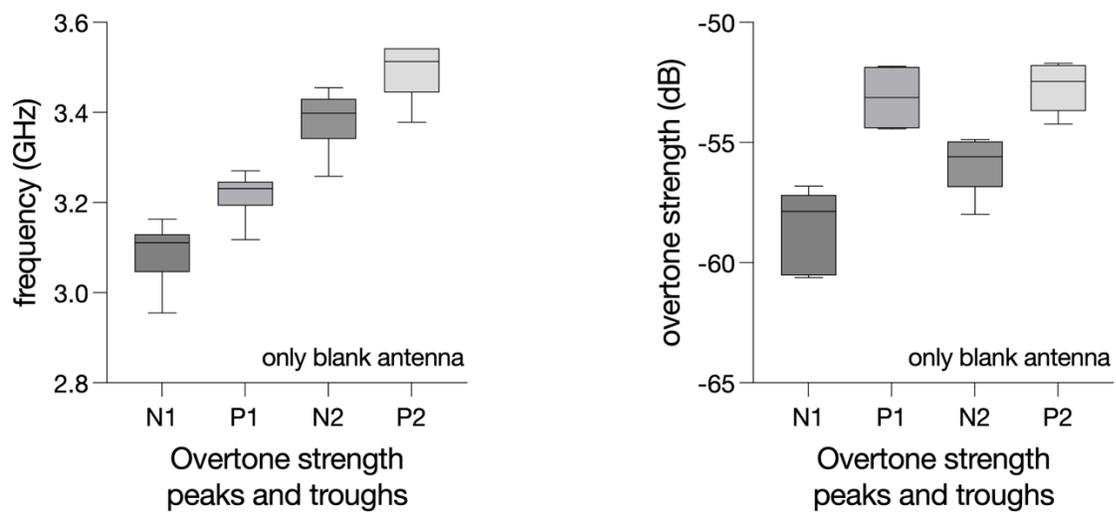

**Fig. S13.** Average frequency and strength of blank µBot overtones.

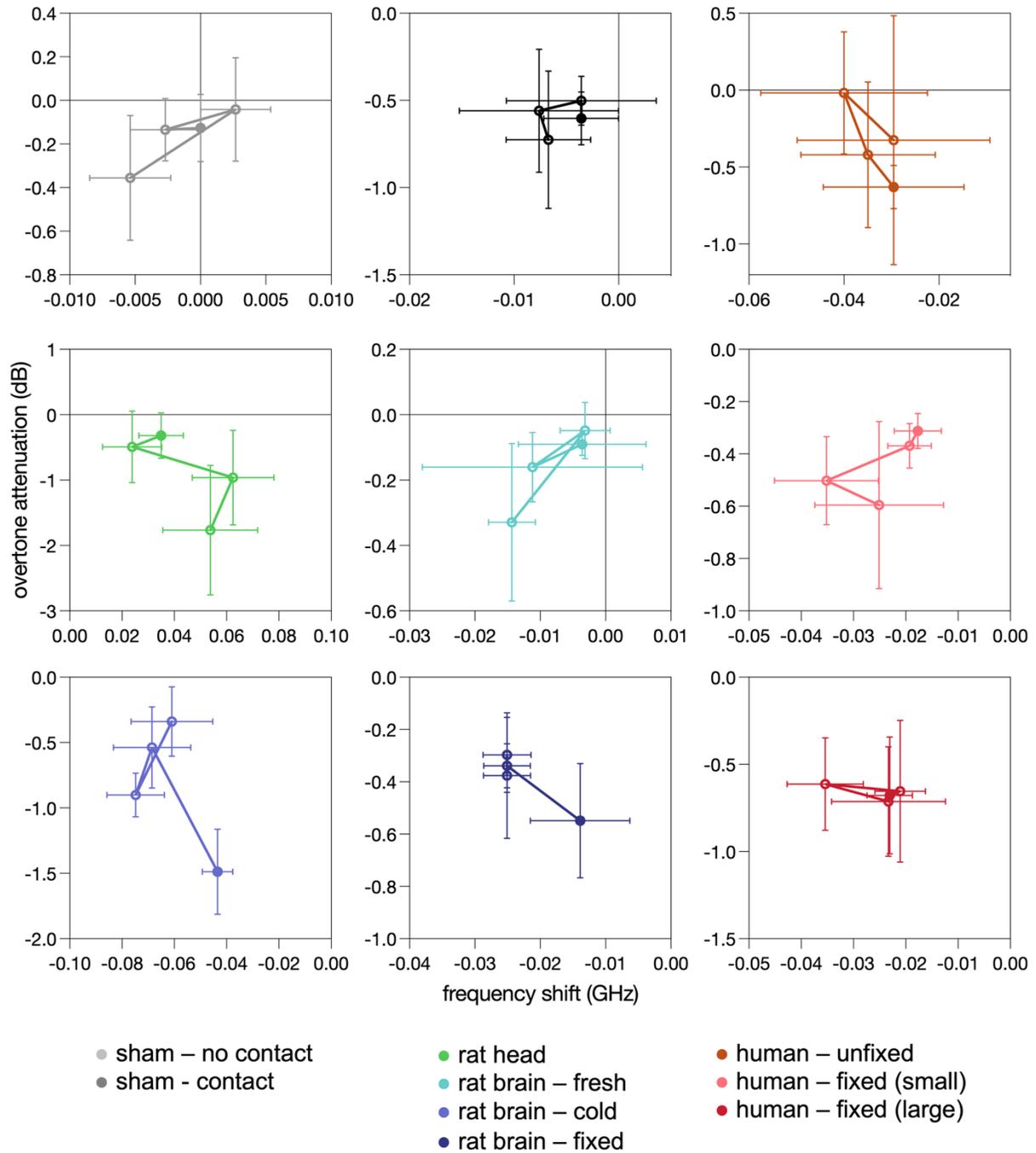

**Fig. S14. Decomposition of each biological tissue impact on the different signal features.** Data presented as mean ± SEM. The filled dot represents N1 and is then sequentially connected to P1, N2, and P2.

SI Note 1: Sonogram video and audio sourced from https://www.youtube.com/watch?v=9Wwu01jdqL8 and converted to *.ts* for data transmission

SI Note 2: CAD file of B205-mini from https://kb.ettus.com/B200/B210/B200mini/B205mini/B206mini

SI Video 1: Calcium imaging on AlN

SI Video 2: Calcium imaging on FeGa

SI Video 3: Calcium imaging on AlN/FeGa/Parylene

SI Video 4: Control calcium imaging on glass slide

SI Video 5-8: MRI Scans post 10-hour implantation of μBots in Agar

SI Video 9: Audio-Visual transmission of sonogram with heartbeat by magnetoelectric antennas as both receiver and transceiver